\documentclass[reprint,superscriptaddress,onecolumn,showpacs,floatfix,longbibliography,nofootinbib]{revtex4-2}

\usepackage{latexsym,amsmath,amsfonts,amssymb,bm,empheq} 
\usepackage{nicefrac} 
\usepackage{esint}
\usepackage{xcolor}
\usepackage{tabularx}  
\usepackage{multirow}  
\usepackage{afterpage}
\usepackage{comment}
\usepackage{subcaption}

\usepackage{tikz}
\usetikzlibrary{arrows.meta,decorations.markings,bending,shapes.misc,calc}
\tikzset{cross/.style={cross out, draw=black, minimum size=2*(#1-\pgflinewidth), inner sep=0pt, outer sep=0pt}, cross/.default={3.5pt}}

\definecolor{ceruleanblue}{rgb}{0.16, 0.32, 0.75}
\definecolor{coolblack}{rgb}{0.07, 0.18, 0.35}
\definecolor{brandeisblue}{rgb}{0.0, 0.44, 1.0}
\definecolor{blue(pigment)}{rgb}{0.2, 0.2, 0.6}
\definecolor{brightmaroon}{rgb}{0.76, 0.13, 0.28}
\definecolor{cobalt}{rgb}{0.0, 0.28, 0.67}
\definecolor{darkblue}{rgb}{0.0, 0.0, 0.55}
\definecolor{darkmidnightblue}{rgb}{0.0, 0.2, 0.4}
\definecolor{dukeblue}{rgb}{0.0, 0.0, 0.61}
\definecolor{indigo(dye)}{rgb}{0.0, 0.25, 0.42}
\definecolor{midnightblue}{rgb}{0.1, 0.1, 0.44}
\definecolor{oxfordblue}{rgb}{0.0, 0.13, 0.28}

\usepackage{mathtools}

\definecolor{nblue}{RGB}{28,130,185}
\definecolor{cgreen}{RGB}{76,153,0}
\definecolor{myorange}{RGB}{245,156,74}

\usepackage{hyperref}
\hypersetup{colorlinks=true, citecolor=magenta, urlcolor=-myorange}

\newcommand{\dif}{\mathrm{d}} 

\newcommand{\calF}{\mathcal{F}}
\newcommand{\calD}{\mathcal{D}}

\newcommand{\calE}{\mathcal{E}} 

\newcommand{\calA}{\mathcal{A}} 
\newcommand{\calAs}{\mathcal{A}_{\mathrm{s}}} 

\newcommand{\ellb}{\ell_{\rm B} } 

\newcommand{\expval}[1]{\langle {#1} \rangle}
\newcommand{\eqexpval}[1]{\langle {#1} \rangle_{\mathrm{eq}}}

\newcommand{\ctwobulk}{c^{(2)}_{\mathrm{bulk}}}

\newcommand{\epsin}{\epsilon_{\mathrm{in}}}
\newcommand{\epsout}{\epsilon_{\mathrm{out}}}

\newcommand{\tauE}{\vartheta_{_\calE}}

\newcommand{\KS}{Keller--Segel }

\newcommand{\JKS}{\bm{J}_{\mathrm{KS}}}

\newcommand{\calFKS}{\mathcal{F}_{\mathrm{KS}}}
\newcommand{\jD}{\bm{J}_{\mathrm{D}}}

\newcolumntype{P}[1]{>{\centering\arraybackslash}p{#1}}

\begin{document}

\title{Nonequilibrium Phenomena in Driven and Active Coulomb Field Theories}
%
\author{Saeed Mahdisoltani}
\email{saeedmah@mit.edu}
\affiliation{Institute for Medical Engineering and Science, Department of Chemical Engineering, and Department of Physics, Massachusetts Institute of Technology, Cambridge, Massachusetts 02139, USA}
\affiliation{Max Planck Institute for Dynamics and Self-Organization (MPIDS), D-37077 G\"ottingen, Germany}
%
%
\author{Ramin Golestanian}
\email{ramin.golestanian@ds.mpg.de}
\affiliation{Max Planck Institute for Dynamics and Self-Organization (MPIDS), D-37077 G\"ottingen, Germany}
\affiliation{Rudolf Peierls Centre for Theoretical Physics, University of Oxford, Oxford OX1 3PU, United Kingdom}

\begin{abstract}
The classical Coulomb gas model has served as one of the most versatile frameworks in statistical physics, connecting a vast range of phenomena across many different areas. Nonequilibrium generalisations of this model have so far been studied much more scarcely. With the abundance of contemporary research into active and driven systems, one would naturally expect that such generalisations of systems with long-ranged Coulomb-like interactions will form a fertile playground for interesting developments. Here, we present two examples of novel macroscopic behaviour that arise from nonequilibrium fluctuations in long-range interacting systems, namely (1) unscreened long-ranged correlations in strong electrolytes driven by an external electric field and the associated fluctuation-induced forces in the  confined Casimir geometry, and (2) out-of-equilibrium critical behaviour in self-chemotactic models that incorporate the particle polarity in the chemotactic response of the cells.  
Both of these systems have nonlocal Coulomb-like  interactions among their constituent particles, namely, the electrostatic interactions in the case of the driven electrolyte, and the chemotactic forces mediated by fast-diffusing signals in the case of self-chemotactic systems. 
The results presented here hint to the rich phenomenology of nonequilibrium effects that can arise from strong fluctuations in Coulomb interacting systems, and a rich variety of potential future directions, which are discussed.
\end{abstract}
\maketitle

\section{Introduction}

Long-ranged Coulomb interactions are prevalent in many systems such as gravitational and electrostatic systems, plasmas, electrolytes, two-dimensional vortices, and chemotactic cell colonies~\cite{LRI-birdview,chavanis-LRI-V}. 
%
Systems that consist of Coulomb-interacting units have deep roots in statistical physics and have been studied for a long time.   
In fact, theoretical investigations of the physical properties of electrolyte solutions marks one of the first applications of statistical mechanics in the nineteenth and twentieth centuries~\cite{levin2002correlations,wrightelectrolyte}.

Systematic efforts in this direction can be traced back to the mean-field model proposed by Arrhenius,  in which an electrolyte solution was modelled as an ideal gas of ionised salts~\cite{arrhenius};   
The failure of Arrhenius's model in explaining the properties of strong electrolytes then  triggered a body of fruitful research that culminated in the celebrated Debye--H\"{u}ckel (DH) theory~\cite{debye1923}. 
Since the introduction of the DH theory, Coulomb systems have been a central part of the statistical physics literature. 
For instance, they have been essential in understanding the properties of two-dimensional systems in the context of Kosterlitz--Thouless physics, covering  such important 
phase transitions as that of the superfluid and  superconductor~\cite{kosterlitz,kosterlitz-review}. 
The concepts and phenomenology of the associated  topological defects have also been utilised in studying the rich collective physics of living systems such as cells and bacteria.  The latter systems are broadly categorised as `active matter', indicating that they consist of particles that move around and exert forces on their surrounding by energy consumption or conversion, as opposed to the traditional `passive' particles~\cite{marchetti2013,ramaswamy-rev}. 
It is worth mentioning that the Poisson form that underlies the long-range Coulomb interactions also governs  the long-time limit of a diffusion processes with source terms~\cite{barton1989}. 
This feature is particularly pertinent to  systems of living or synthetic entities whose interactions are typically mediated by chemical signalling molecules that diffuse in the environment. 
In the relevant regime of fast-diffusing signals, the chemical concentration is governed by a Poisson-like relation, indicative of Coulomb-like interactions among the constituent entities~\cite{golestanian2012,tsori2004,chavanis2008,golestanian2009anomalous,chavanis-jeans-2008,gelimson2015,agudo-activephase2019,golestanian2019}.

%
%
In equilibrium conditions, the powerful screening picture of the DH theory introduces a characteristic `Debye screening length', beyond which correlations in electrostatic systems die out exponentially with distance. 
As a result, in the presence of screening effects, Coulomb systems at sufficiently large length scales can be considered weakly correlated, and an effective coarse-grained  description of them may be achieved in the spirit of the central limit theorem~\cite{goldenfeld1992,levin2002correlations}. 
Only in the presence of strong fluctuations and correlations, e.g. at a critical point, large-scale distributions will instead be governed by scaling relations; in such cases, to determine the functional form of the emerging distribution, one needs to employ the more sophisticated field-theoretical approach and the  renormalisation group (RG)  theory machinery~\cite{goldenfeld1992,fisher-RG-rev,stanley-RG-rev}. 
%
%
Strong fluctuations and critical-like behaviour are, in fact, ubiquitous out of thermal equilibrium and in dynamical situations. For instance, conservation laws can generically (i.e. with no fine tuning) give rise to long-range correlations in diffusive dynamics~\cite{grinstein91generic,zia-schmittmann}. 
Since realistic systems often operate away from equilibrium conditions, one finds a diverse range of nonequilibrium scaling relations in natural, biological, and synthetic systems~\cite{marchetti2013,vicsek-collective-review,chialvo,sornette-book,BTW,racz-lecturenotes}. 
A general understanding of nonequilibrium phenomena in systems with Coulomb interactions is still lacking in the literature, and the bulk of the existing results rely on various mean-field assumptions~\cite{chavanis-LRI-V,nanofluidsreview}.   
The prospect of novel collective physics in nonequilibrium systems thus motivates the investigation of Coulomb systems out of thermal equilibrium  {and beyond the DH picture}. 
%
%
Here, we review and discuss some recent results on two nonequilibrium systems with long-range interactions, based on Refs.~\cite{mahdisoltani2021long,mahdisoltani-NJP,mahdisoltani2021chemotaxis,zinati-cg2021}. 
In Section~\ref{sec:electro}, the simplest case of a nonequilibrium Coulomb system is considered, namely a strong electrolyte that is driven by a constant external electric field. 
Analysing the stochastic dynamics of the ions in the driven electrolyte reveals that the counterion screening effects -- which are dominant in the equilibrium regime -- are overruled by a dynamical mechanism known as \textit{generic scale invariance}~\cite{tauber}: as soon as the external electric field is switched on, the driven electrolyte becomes  long-range correlated without the need for parameter tuning. 
Apart from its conceptual importance in modelling nonequilibrium charged solutions, such correlations have practical implications in, e.g., force propagation across electrolyte films. 
An example of these implications in the context of fluctuation-induced forces (FIFs) in the flat Casimir geometry, which survive beyond the Debye screening length and have transient regimes that decay algebraically slowly over time, are reviewed. 
In Section~\ref{sec:chemo}, the focus is on the 
the stochastic dynamics of a \textit{self-chemotactic} system, where each particle releases chemical signalling molecules in the environment and responds to the resulting chemical gradients by adjusting its motion. 
Scaling analysis shows how the well-known Keller--Segel (KS) model should be modified by incorporating additional unconventional chemotactic couplings in order for it to capture the correct macroscopic features. 
For the nonequilibrium coupling that is associated with particle-polarity effects in chemotaxis, the symmetry properties allow for an exact evaluation of scaling exponents. 
 %
%
Pertinent points in both cases are discussed at the end of the respective sections, and concluding remarks and potential directions for future studies are discussed in Section~\ref{sec:conc}.

\section{Driven electrolyte}    \label{sec:electro}

%
A strong  electrolyte that consists of an equal number of cations and anions with charges $\pm Q$ is driven out of equilibrium by an external electric field $\bm{E} = E \hat{\bm{e}}_x$. 
For the sake of simplicity, it is assumed for now that ions in the solution have equal mobility factors 
$\mu_+=\mu_- \equiv \mu$, 
although relaxing this assumption will be discussed later. 
 {Note that the hydrodynamic effects of the solvent are neglected throughout this analysis}%
\footnote{ {Hydrodynamic correlations are generally long-ranged and exhibit algebraic decays~\cite{R1-1,R1-2}, and they can become important  in, for instance, concentrated electrolytes, where  the inter-particle distances are comparable with the  hydrodynamic radii of the particles, as well as in electrolytes under extreme confinements. 
As we do not consider concentrated electrolytes, and also since the focus is on the case of large separations between the confining plates (as compared to the Debye screening length of the electrolyte), such hydrodynamic effects are ignored in this  analysis.  
This simplification is also in line with cues from some of the recent force measurement experiments where one observes that the viscosity and hydrodynamic effects of the solvent cannot be responsible for the observed long-range effects;  see, e.g., Ref.~\cite{perez2019surface}. 
Lastly, it is worth mentioning that even in the presence of the long-ranged hydrodynamic correlations, the resulting hydrodynamic contribution to the electrolyte pressure is still found to decay exponentially with distance, see Ref.~\cite{ajdarinote}; 
as such, the hydrodynamic contribution is not expected to affect the forces exerted by the driven electrolyte on the plates in the asymptotic large-separation regime.
}}.
Each positive or negative ion 
$a \in \lbrace 1,2,\ldots,N \rbrace$ in the electrolyte solution is subject to a thermal Gaussian white noise $\bm{\eta}^\pm_a(t)$~\cite{onsagerlong1932} and is also affected by the electric field ($-\nabla \phi$) emanating from other ions in the solution, as well as by the external field $\bm{E}$. 
The Langevin description of the overdamped trajectory $\bm{r}^\pm_a(t)$ of a cation or anion reads~\cite{zorkot2016power,demery2016conductivity}
\begin{equation}       \label{eq:langevinparticle}
    \frac{d{\bm{r}}_a^\pm}{dt} = 
    \pm \mu Q \left[ -\nabla\phi (\bm{r}_a) + \bm{E} \right] + 
    \sqrt { 2 D } \, \bm{\eta}_a^\pm (t),  
\end{equation}    
with noise correlations given by 
$\langle \eta_{a i}^+ (t) \eta_{b j}^+ (t') \rangle = 
\langle \eta_{a i}^- (t) \eta_{b j}^- (t') \rangle =
\delta_{ab} \, \delta_{ij} \, \delta ( t - t' )$
($i$ and $j$ represent vectorial  components in $d$ dimensions).
The noise strength $D$ and the ionic mobility $\mu$ are related through the fluctuation--dissipation relation as $D=\mu k_B T$, where $T$ is the temperature of the electrolyte~\cite{kardar-statfield}.  

In order to represent the Poisson equation that governs the electric potential $\phi$, first the instantaneous number density of cations and anions is defined as 
$ C^\pm(\bm{r},t) = \sum_{a} 
 \delta^d \left( \bm{r} - \bm{r}_a^\pm(t) \right)$; 
the electrostatic Poisson equation then reads 
\begin{align}
 -\nabla^2 \phi = \frac{S_d  Q}{\epsilon_{\rm in}}( C^+ - C^-), 
\end{align} 
 where 
$S_d = \dfrac{2 \pi^{d/2}}{\Gamma(d/2)}$ is the area of the  $d$-dimensional sphere with unit radius and $\epsin$ is the permittivity of the solution. 
The Dean--Kawasaki (DK) formalism \cite{dean96langevin,kawasaki1994,frusawa2022electric,ddftreview} then directly yields the exact dynamics of $C^\pm$ in the form of stochastic continuity equations~\cite{mahdisoltani2021long,mahdisoltani-NJP}. 

The dynamics of density and charge fluctuations around a uniform state of the electrolyte are described by the linearised DK equations.  
These are obtained by rewriting the full equations using 
$C^\pm = C_0 + \delta C^\pm$,   
where $C_0$ is the equal uniform background density of both cations and anions and $\delta C^\pm$ are their  local fluctuations.  
These fluctuations are assumed to remain small compared to the background density, i.e. $\delta C^\pm \ll C_0$. 
Accordingly, the fluctuations in the total number of both cations and anions are 
$c (\bm{r},t) = \delta C^+  +  \delta C^-$ 
while the corresponding charge fluctuations read 
$\rho (\bm{r},t) = \delta C^+  -  \delta C^-$ (in units of $Q$).  
The resulting linearised DK equations  read~\cite{mahdisoltani2021long,mahdisoltani-NJP}
\begin{align}   
    \partial_t c &= 
    D\nabla^2 c - \mu  Q  \bm{E}\cdot\nabla \rho + \sqrt{4DC_0} \, \eta_c,  \label{eq:linc} \\
    \partial_t \rho &=
    D\nabla^2 \rho - \mu  Q\bm{E}\cdot \nabla c - D \kappa^2 \rho + \sqrt{4DC_0} \, \eta_\rho     \label{eq:linrho}, 
\end{align}
where 
$\kappa = \sqrt{2 S_d C_0 \ellb}$
with the `Bjerrum length'  
$\ellb = \beta  Q^2 / \epsilon_{\rm in}$; the Debye screening length is then given by $\kappa^{-1}$. 
Moreover, the additive noise terms $\eta_{\rho}$ and $\eta_c$ have zero means and are characterised by the correlations 
$\langle \eta_\rho (\bm{r},t) \eta_\rho (\bm{r}',t') \rangle = \langle \eta_c (\bm{r},t) \eta_c (\bm{r}',t') \rangle = 
-\nabla^2 \delta^d (\bm{r}-\bm{r}')  \delta (t-t') $.

Standard scaling analysis of the nonlinear interaction terms in the full DK equation before linearisation  shows that close to the Gaussian fixed point, all such terms are irrelevant at the macroscopic level~\cite{mahdisoltani-NJP}. 
It is straightforward to show that the multiplicative noise contributions are also irrelevant in the macroscopic limit. 
Therefore, the linearized stochastic equations~\eqref{eq:linc} and \eqref{eq:linrho} are sufficient to capture the asymptotic large-scale behaviour of the ionic  dynamics.  
In this limit, the density dynamics~\eqref{eq:linc} constitutes the `slow' (i.e. diffusive) process in the system, whereas the charge dynamics~\eqref{eq:linrho} is the `fast' (i.e. relaxational) mode. 

The distinction is already evident for the non-driven  $\bm{E}=0$ case, in which the diffusive density fluctuations propagate as    
$\propto e^{-t/\vartheta_c(\bm{k})}$    
with a wavevector-dependent relaxation time 
$\vartheta_c (k) = 1/(D k^2)$ 
that diverges for $k\to 0$, while the relaxational charge correlations are  
$ \propto e^{-t/\vartheta_\rho(k)}$,   
with a relaxation time 
$\vartheta_\rho(k) = 1/\left[D(\kappa^2 + k^2)\right]$ that has a finite value $1/D\kappa^2$ in the  long wavelength limit.  

Furthermore, it can be observed that for $\bm{E}=0$, the dynamics of the density and charge fluctuations are decoupled.  However,  switching on the external field 
$\bm{E}$ couples these dynamics through the terms
$\propto E \partial_x \rho$ 
and 
$\propto E \partial_x c$ in the linearised equations.  
Because of the resulting interdependence between density and charge dynamics, and since the density fluctuations comprise the slow diffusive mode, the distribution of the charge fluctuations has to quickly adapt to variations in the density distribution. 
Indeed, the scaling analysis suggests that for the asymptotic large-scale limit, the leading form of the charge profile~\eqref{eq:linrho} is given by~\cite{mahdisoltani2021long}
\begin{align} \label{eq:rho(c)}
    \rho (\bm{r},t) \approx -\kappa^{-2}\beta Q E \partial_x c(\bm{r},t).    
\end{align}
We refer to this result as the \textit{quasi-stationary approximation}, since it describes the macroscopic charge fluctuations at scales beyond the Debye screening length $\kappa^{-1}$ and the Debye relaxation time $\vartheta_\rho (k=0)$.  
Note that at thermal equilibrium, charge fluctuations do not survive at these macroscopic scales, in agreement with the fact that $\rho$ in  Eq.~\eqref{eq:rho(c)}  vanishes for $E=0$. %

Apart from the scaling analysis, one may intuitively understand Eq.~\eqref{eq:rho(c)} through the following picture: 
the presence of the external field perturbs the so-called `counterion atmospheres' that surround each ion in the electrolyte~\cite{onsagerlong1932}. 
In a uniform distribution of the ions, charge imbalances due to these  disturbances cancel out each other; however, in non-uniform distributions  where there is a density gradient parallel to the external field, the accumulation of charge imbalance can lead to an effective large-scale charge profile described by Eq.~\eqref{eq:rho(c)}. 
%

The effective dynamics of the density fluctuations $c$ in this asymptotic limit is consequently obtained by substituting the  quasi-stationary charge profile Eq.~\eqref{eq:rho(c)} into Eq.~\eqref{eq:linc}. 
One thus arrives at the \textit{anisotropic diffusion} equation\footnote{The noise term $\eta_\rho$ is discarded from Eq.~\eqref{eq:rho(c)} since its contribution is irrelevant with respect to  that of $\eta_c$ in Eq.~\eqref{eq:anisodiffusion}.} 
\begin{align}   \label{eq:anisodiffusion}
    \partial_t c = D (\calE^2 \partial_x^2 + \nabla^2)c + \sqrt{4DC_0} \, \eta_c, 
    \qquad \text{with} \qquad 
    \calE = \frac{\mu Q E}{D\kappa} = \left[ \frac{\epsin E^2/(2S_d)}{C_0 k_B T} \right]^{1/2}.
\end{align}
%
%
%
The dimensionless electric field $\calE$ can be understood through the counterion atmosphere as follows: 
the disturbance in each atmosphere is characterised by its relative asymmetry in the direction of the external field, which is determined by the distance the central ion moves during the relaxation period of the counterion cloud. 
This distance is given by the product of an ion's deterministic speed,   
$\mu Q E$, 
and the relaxation time of the cloud, 
$1/(D\kappa^2)$; 
calculating the resulting displacement, and dividing it by the spatial spread of the cloud ($\sim \kappa^{-1}$), one arrives at the desired expression for $\calE$~\cite{mahdisoltani-NJP,onsagerlong1932}. 

Equation~\eqref{eq:anisodiffusion} indicates  that the presence of the external electric field enhances the density diffusion along its direction and thus breaks the isotropy of the  conserving density dynamics. 
The resulting anisotropy is not reducible by a rescaling of space and time, due to the fact that the effective noise strength remains the same in all directions. 
The associated imbalance between the dissipative ionic current and the noisy fluctuations lead to the so-called \textit{generic scale-invariant}~\cite{grinstein90conservation,grinstein91generic,zia-schmittmann,tauber}: the correlations take the form of power-law functions in the general case without any parameter tuning.    

A standard calculation yields these density correlation functions from Eq.~\eqref{eq:anisodiffusion} as
%
$\expval{c(\bm{k},t) c(\bm{k}',t')} = \eqexpval{c(\bm{k},t) c(\bm{k}',t')} + (2\pi)^d \delta^d(\bm{k}+\bm{k}') \, \ctwobulk (\bm{k},t ,t'),$
where the averaging is taken with respect to both the noise realisations and thermal initial configurations~\cite{mahdisoltani-NJP},
and the `equilibrium-like' contribution is the diffusive, short-ranged contribution that recovers the equilibrium correlations for $\bm{E}=0$ and reads  
$\eqexpval{c(\bm{k},t) c(\bm{k}',t')} = (2\pi)^d \delta^d (\bm{k}+\bm{k}') \, 2C_0 e^{-D (\calE^2 k_x^2 + k^2) (t'-t)}$. 
The nonequilibrium electric-field-dependent contribution, on the other hand, is given by  (assuming $t'\geq t$)
\begin{equation}    \label{eq:c2bulk-Fourier}
\begin{split}
    & \ctwobulk (\bm{k}, t, t')=  
     \frac{ - 2 C_0 \calE^2 k_x^2}{\calE^2 k_x^2 + k^2}
    \left[ \exp\left(-\frac{t'-t}{\tauE(k)} \right) - \exp\left(-\frac{t'+t}{\tauE(k)} \right) \right], 
    \qquad \text{with} \qquad 
    \tauE (k) = \frac{1}{D(\calE^2 k_x^2 + k^2)}, 
\end{split}    
\end{equation}
which clearly represents correlations caused by the driving field. 
The charge correlation functions subsequently follow from Eq.~\eqref{eq:c2bulk-Fourier} by taking appropriate derivatives according to the quasi-stationary relation~\eqref{eq:rho(c)}.

The Fourier representation of the nonequilibrium density correlation~\eqref{eq:c2bulk-Fourier} has an anomalous `discontinuity singularity' in the long-wavelength limit ($k\to 0$)~\cite{tauber,zia-schmittmann}, which is indicative of its power-law decay with distance in the real space. 
For example, for $t=t'$, 
 {one can introduce a new wavevector $\tilde{\bm{k}}$ whose components are the same as $\bm{k}$ except that $\tilde{k}_x = k_x \sqrt{\calE^2+1}$. 
Then by making use of the convolution theorem,  
transforming Eq.~\eqref{eq:c2bulk-Fourier} back to the real space gives}
\begin{equation}   \label{eq:c2bulk-real}
\begin{split}
    \ctwobulk (\bm{r},t) = 
    &-\frac{2C_0 \calE^2 (1-d  \tilde{x}^2/\tilde{r}^2)}{ S_d(\calE^2+1)^{3/2} \tilde{r}^d}
    +  \int 
    \dif^d \tilde{\bm{r}}' \, 
    \frac{2C_0 \calE^2 (1-d  {\tilde{x}}^{\prime 2}/\tilde{r}^{\prime 2})}{ S_d(\calE^2+1)^2 \,  \tilde{r}^{\prime d}} \,
   \frac{\exp\left(-\frac{(\tilde{\bm{r}}-\tilde{\bm{r}}')^2 }{8 D t}\right)}{  (8\pi D t)^{d/2}  }, 
\end{split}    
\end{equation}
where
 {the second term on the r.h.s. is the convolution of the long-range correlation function with the diffusion kernel, and 
all components of $\tilde{\bm{r}}$ are the same as $\bm{r}$ except that $\tilde{x} = x / \sqrt{\calE^2+1}$.}
Equation~\eqref{eq:c2bulk-real} clearly shows that in $d$ dimensions, the density correlation function falls off with distance as $\sim r^{-d}$, and it has a dipolar form typically seen in driven systems~\cite{tauber,zia-schmittmann}. 
The transient density correlations, encapsulated by the second term on the r.h.s. of Eq.~\eqref{eq:c2bulk-real},  display similar spatial features but in a temporally transient way, and their long-time tail behaves as $\sim t^{-d/2}$. 
Since the real-space form of the charge correlations is also obtained from Eq.~\eqref{eq:c2bulk-real} upon  taking additional derivatives, it becomes evident that the nonequilibrium charge fluctuations are long-range correlated as well, despite the Debye screening effects.  
%

%
\subsection{Long-ranged fluctuation-induced force in driven electrolytes}

\begin{figure}[t]
		(a)	\begin{minipage}[c]{.4\linewidth}
		\centering
		\includegraphics[width=\linewidth]{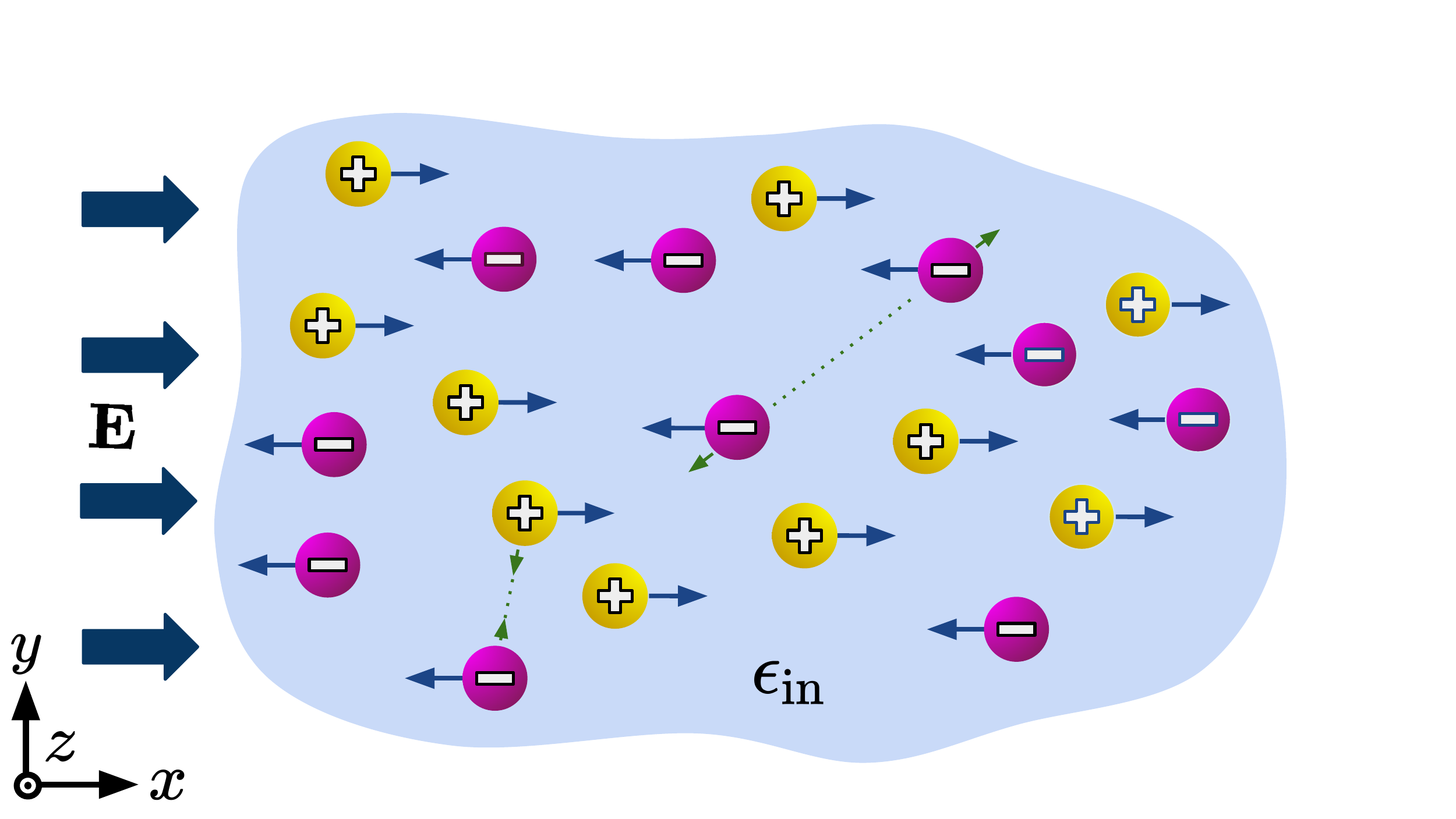}
	        \end{minipage} 
	\hskip.5cm
        (b)	\begin{minipage}[c]{.4\linewidth}
		\centering
		\includegraphics[width=\linewidth]{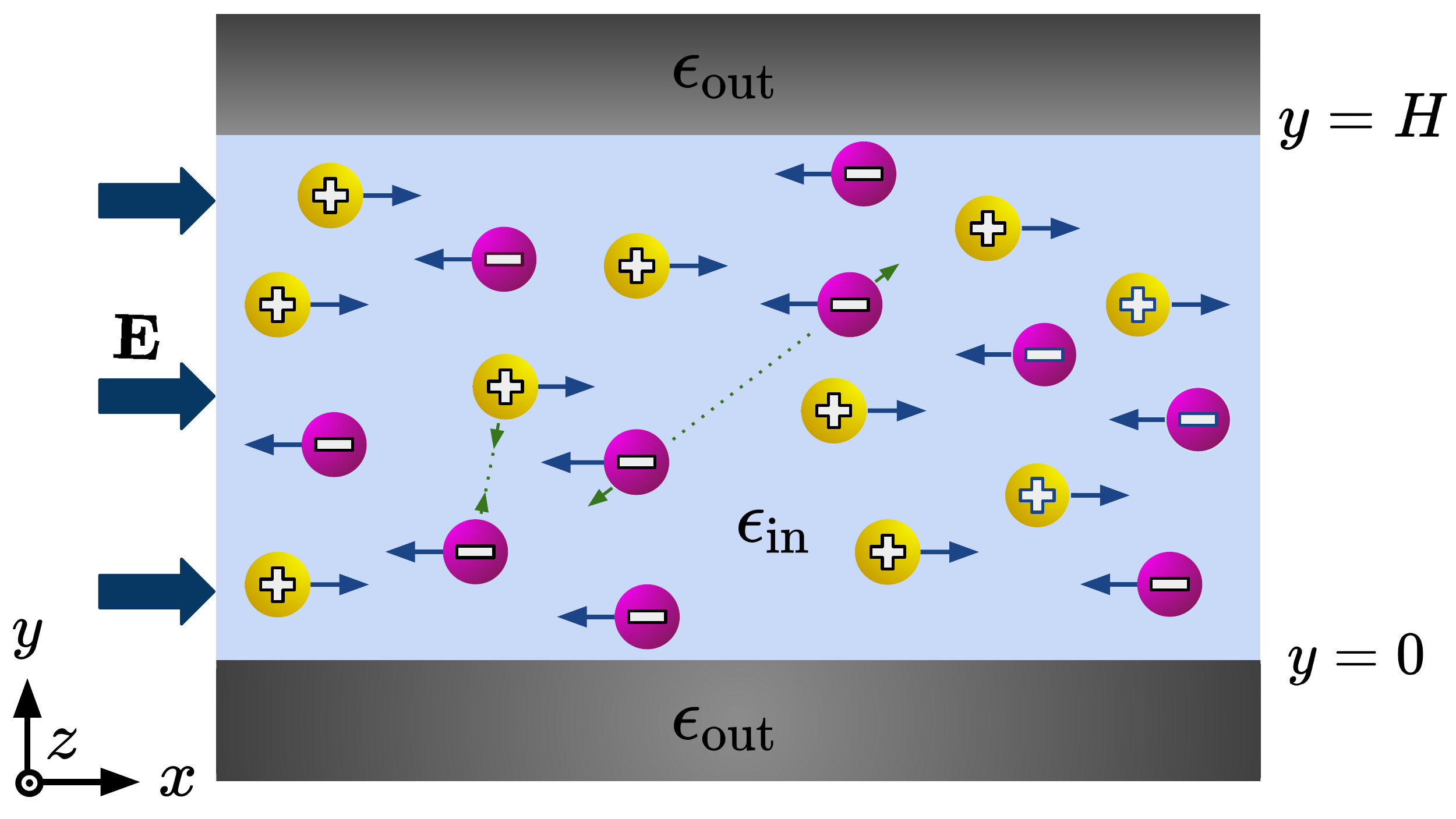}
		    \end{minipage} 
	\caption{ The $3$-dimensional case of a driven electrolyte in (a) the bulk and in (b) the confined flat Casimir geometry.  
	The external electric field $\bm{E}=E \hat{\bm{e}}_x$ is switched on at $t=0$ and drives the cations and anions in opposite directions (blue arrows), while the ions exert electrostatic forces on each other (green arrows). 
	The symmetric confining plates in panel (b) are electrically uncharged and confine the electrolyte in the $y$ direction between $y=0$ and $y=H$, while the $x$ and $z$ directions remain open.
	} \label{fig:schematic-confined}
\end{figure}

The long-range nature of the correlation functions in a driven electrolyte has direct implications in terms of force propagation across the electrolyte. 
In particular, it is well-known that confining  long-ranged fluctuations gives rise to \textit{fluctuation-induced forces} (FIFs) resembling the original quantum Casimir effect~\cite{kardar99friction,gambassi2009review,fisher1978wall} which can act across large separations beyond the Debye length in the present case (typically $\kappa^{-1} \sim 1-10\,\mathrm{nm}$). 
The FIF arising from confining a driven electrolyte film between two parallel plates is reviewed below  (Fig.~\ref{fig:schematic-confined}). 

The driven electrolyte exerts a pressure on the boundary plates that can be calculated using the normal component of the noise-averaged  Maxwell stress tensor~\footnote{The shear components vanish due to symmetry.}, namely 
\begin{align}   \label{eq:maxwell-stress-yy-avg}
    \expval{\sigma_{yy}} = \frac{\epsin}{2 S_d} 
    \Big( \left\langle \left(\partial_y \phi\right)^2 \right\rangle -
    \left\langle \left(\nabla_{\bm{s}} \phi\right)^2 \right\rangle  \Big), 
\end{align}
where $\bm{r} = (y, \bm{s})$ and the parallel components are  
$\bm{s} = (s_1 \!=\! x, s_2,\ldots,s_{d-1}) \in \mathbb{R}^{d-1}$. 
Evaluating this expression requires the correlations of the electric field created by the ions in the solution. 
To obtain these, the electric field can be expressed in terms of the charge density $\rho$ by making use of  the method of image charges~\cite{jackson}, and therefore the associated field correlations can be rewritten in terms of the charge correlations in confinement. 
The latter correlations are obtained similar to the bulk case discussed in the previous subsection; however, to satisfy the no-flux boundary conditions imposed by the plates, one needs to decompose the fluctuation fields $c$ and $\rho$ onto the Neumann  eigenfunctions. 
On performing this calculation, one eventually arrives at the following expression for the net force per unit area acting on the $y=H$~\footnote{The forces exerted on the plates are equal and in opposite directions. 
 {Note that the kinetic (ideal gas) forces exerted on either side of each plate are equal in magnitude and thus do not contribute to the net force.}} boundary~\cite{mahdisoltani-NJP} :
\begin{align}   \label{eq:FoverS(t)}
    \frac{F(t)}{S} = -\frac{k_B T}{H^d} \calE^4 \, \calA (\calE,\lambda,t),  
\end{align}
where the `dielectric contrast' between the solvent and the boundaries determines the extent of the charge  polarisation in the boundaries and is given by 
$\lambda = \dfrac{\epsin - \epsout}{\epsin+ \epsout}$. 
Equation~\eqref{eq:FoverS(t)} represents an attractive FIF between the plates if the dimensionless amplitude $\calA$ is positive,  and a repulsive one if it is negative. 

Through standard integration techniques, one can obtain some explicit expressions for both the transient and the steady-state parts of the dimensionless amplitude $\calA$; these are rather lengthy and can be found in Refs.~\cite{mahdisoltani2021long, mahdisoltani-NJP}. 
From Eq.~\eqref{eq:FoverS(t)}, it is already evident that the FIF scales as $\sim H^{-d}$ with plate separation. 
The temporal variations of the transient part of the FIF, on the other hand, is complicated and requires a more detailed examination; in the long-time limit, such contributions generally decay algebraically slowly with time as $\sim t^{-d/2}$, except for $\lambda \lesssim 1$ where the asymptotic limit is given by $\sim t^{-(d-1)/2}$. 
One can additionally deduce that the FIF scales as $\calE^4$ for $\calE\ll 1$ and as $\calE^2$ for $\calE \gg 1$. 
Asymptotic expressions and numerical computations also reveal that $\calA$ can become negative for $\lambda \ll 1$, suggesting a possible regime for achieving repulsive FIF with symmetric boundary conditions.

\begin{figure*}[t]
	(a) \quad	\begin{minipage}[c]{.38\linewidth}
		\centering
		\includegraphics[width=\linewidth]{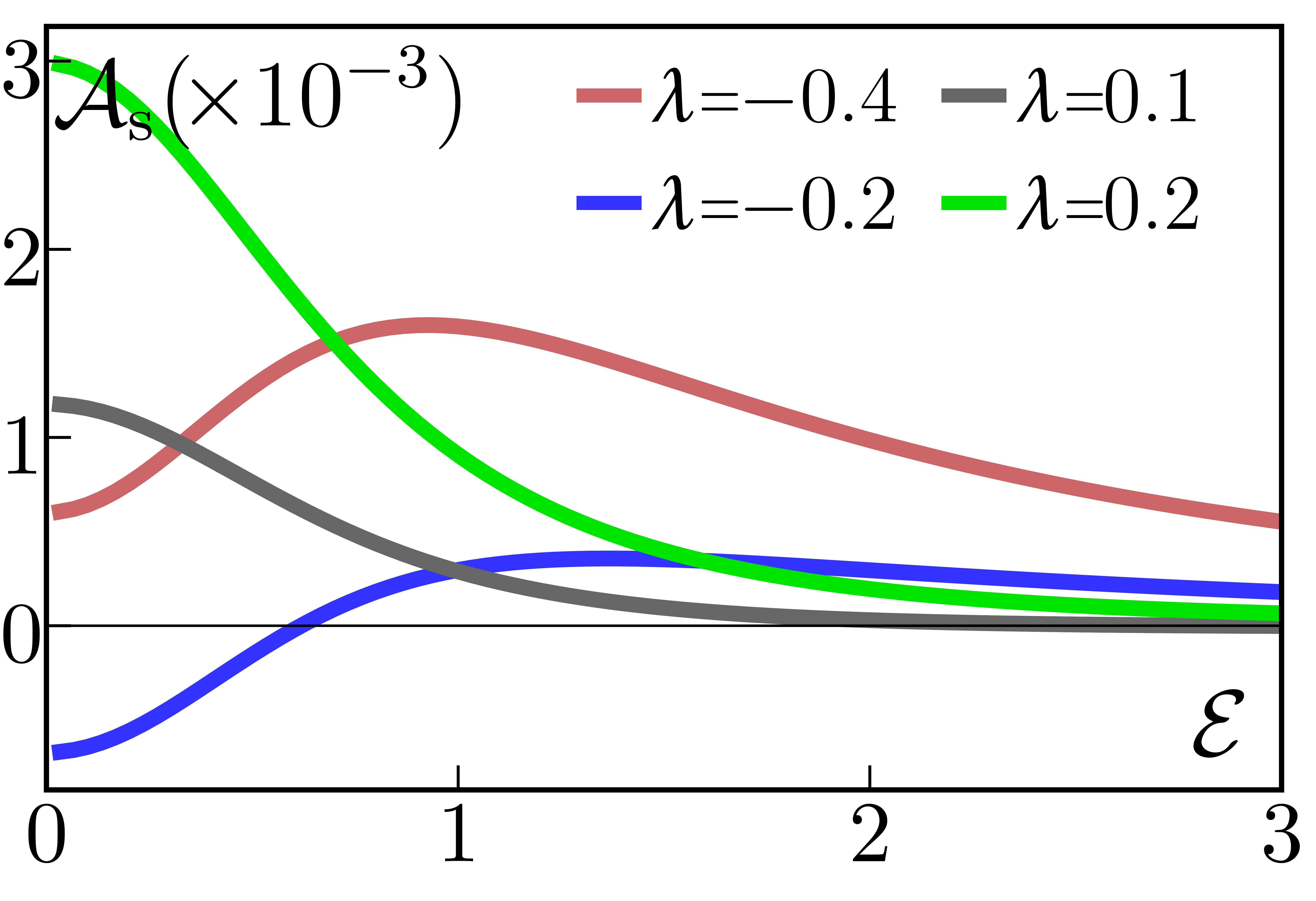}
	        \end{minipage} 
	\hskip.5cm
        (b)	\quad \begin{minipage}[c]{.4\linewidth}
		\centering
		\includegraphics[width=\linewidth]{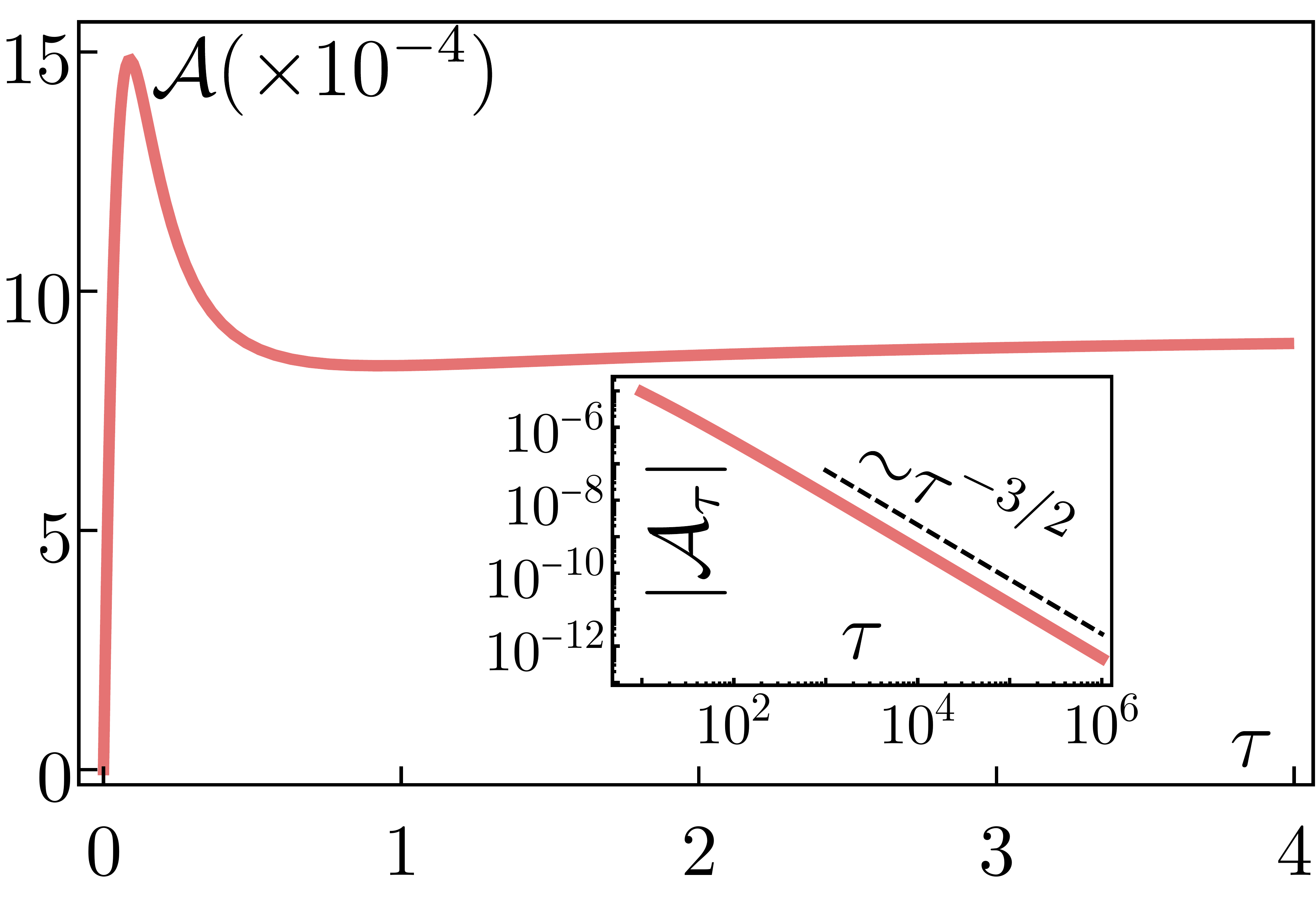}
		    \end{minipage}
	\caption{(a) Variations of the (numerically evaluated) steady-state part of the FIF amplitude, denoted by $\calAs$, with the electric field strength $\calE$. 
	(b) The total FIF amplitude $\calA$ plotted as a function of the dimensionless time variable $\tau = \frac{D t}{H^2}$ (the external field is switched on at $t=0$), for $\lambda=-0.4$ and $\calE=0.3$.   
    The inset displays the  transient part of the force amplitude,  denoted by $\calA_\tau$,  as a function of time in logarithmic scales. 
	} \label{fig:calA_plots}
\end{figure*}


\subsection{Remarks}

Long-range correlations have also been uncovered in the context of driven binary mixtures with short-range interactions~\cite{poncet2017universal}. 
In that case, the universal correlation functions in $d$ dimensions have algebraic spatial decays as $\sim r^{-(d+1)/2}$  only along the direction of the driving force,  
while in the transverse plane correlations still decay exponentially fast. 
Correlation functions of driven electrolytes as considered here, on the other hand, are long-ranged in all directions ($\sim r^{-d/2}$) and they have a dipolar character. 

It is worth mentioning that even though the equations of this section rely on the  assumption that the external electric field is constant in time, in fact they also capture the macroscopic behaviour of the driven electrolyte when $\bm{E}$ varies slowly in time.  
%
For a time-dependent field 
$\bm{E}(t) = E(t) \hat{\bm{e}}_x$, the deterministic part of Eq.~\eqref{eq:linrho} is solved in the Fourier space by 
\begin{align}   \label{eq:rhokt-E(t)}
    \rho(\bm{k},t) = -i \mu Q k_x \int_0^t \dif t' \, \exp\left(-\frac{t-t'}{\vartheta_\rho(k)}\right) \, 
    E(t') \, c(\bm{k},t'), 
\end{align}
where, for simplicity, we have set  
$\rho(\bm{k},t\!=\!0) = 0$. 
Through an appropriate Taylor expansion, it is straightforward to show that the asymptotic long-time limit of this integral expression, when the characteristic time-scale of the electric field is much slower than the Debye time $1/(D\kappa^2)$, is governed by  
$\rho(\bm{k},t) \approx 
    \dfrac{-i \mu Q k_x}{D(\kappa^2 + k^2)} \, 
    E(t) \, c(\bm{k},t).$ 
In the long-distance limit ($\kappa \gg k$), this indeed recovers the Fourier representation of the quasi-stationary relation~\eqref{eq:rho(c)}. 
In other words, if the applied electric field changes sufficiently slowly over time, i.e. its characteristic frequencies are smaller than $D\kappa^2$ (as is the case, e.g., in the experimental settings in Refs.~\cite{perez2019surface,stoneAC}), to the leading approximation the relaxation dynamics of the counterion clouds are unaffected by changes in the external field.  

Lastly, it is noteworthy that the equality of the mobility factors of the cations and anions is also not necessary for the mergence of these long-range correlations.  
In general, charge species in an electrolyte solution have different diffusion coefficients, e.g., due to hydration effects of the solvent~\cite{israelachvili}, and this difference can have important effects on the time-averaged electric fields within the bulk of the electrolyte~\cite{amrei2018oscillating}.
It is straightforward to show that, once an unscreened external electric field is maintained in the bulk of the electrolyte, the difference between the ionic mobilities only adds terms to the model that are irrelevant in the long-distance limit~\cite{mahdisoltani-NJP}. 
Therefore, the effective large-scale description of an electrolyte with unequal ionic mobilities is also given by the same anisotropic diffusion equation~\eqref{eq:anisodiffusion}\footnote{However, in that case  $D$ should be replaced by the average of the diffusion coefficients of the cations and anions~\cite{mahdisoltani-NJP}.}.

\section{Nonequilibrium polarity-induced chemotaxis} 
\label{sec:chemo}

%
Chemotaxis is a prevalent mechanism among living cells through which they sense variations in the concentration of chemical signalling molecules and respond to them by moving up or down the corresponding  gradient~\cite{levine2013,camley2018collective}. 
Prokaryotes such as \textit{Escherichia coli}, with a typical cell size of $\sim 1 \, \mu\mathrm{m}$, often perform gradient measurements while in motion by making temporal comparisons of the concentration over temporal window of a few seconds; 
eukaryotic cells such as neutrophils, on the other hand, are larger in terms of size ($\sim 10\, \mu\mathrm{m}$) and can thus perform a spatial comparison of the signal concentration across their bodies~\cite{levine2013,chemotaxis-glance}.  
Despite the variations in the internal processes that underlie chemotaxis in different cell types~\cite{berg2000motile,levine2013,van2004chemotaxis}, the gradient sensing response of the cell emerges rather generically, and it is present even in enzymes and synthetic active particles~\cite{dey14,agudo2018,Kapral2012,Soto2014,illien2017,stark2018}. 

Chemotactic effects are crucial for a range of population-level  phenomena in living systems, including collective cell migration~\cite{friedl2009,giniunaite2020modelling}, 
morphogenesis~\cite{hogan1999,crick1970}, 
immune response~\cite{wound-review,petri2018neutrophil,fibroblast-wound}, and cancer metastasis~\cite{roussos2011,hanahan2011}. 
Since such collective phenomena appear in various contexts where the cells and the details of their interactions are different, it is plausible to construct effective models that take only a few microscopic details and predict and describe  their qualitative large-scale features~\cite{camley2018collective}. 
An important example of such theoretical models was investigated by Keller and Segel nearly fifty years ago~\cite{keller1971,patlak1953random}. 
%
The so-called Keller--Segel (KS) model describes chemotaxis as a bias in the stochastic trajectory of a moving cell caused by chemical gradients.   
More precisely, in a spatio-temporal chemical field $\Phi$, the overdamped dynamics of a single chemotactic cell is described by 
\begin{align}   \label{eq:KS-Langevin-single}
    \frac{\dif \bm{r}}{\dif t} = \bm{v}_{\mathrm{KS}} + \bm{\xi}(t), 
\end{align}
where $\bm{v}_{\mathrm{KS}} = \nu_1 \nabla\Phi$ is the KS chemotactic drift term, with $\nu_1$ representing the associated mobility coefficient, and the components of the Gaussian white noise $\bm{\xi}$ have correlations  
$\expval{\xi_i (t) \xi_j (t')} = 2 D \delta_{ij} \delta(t-t')$. 

One can use the DK approach to construct the exact stochastic dynamics of the cell density 
$C(\bm{r},t) = \sum\limits_{a=1}^N \delta^d (\bm{r}-\bm{r}_a(t))$ in $d$ dimensions in the form of a continuity equation~\cite{chavanis2010stochasticKS,mahdisoltani2021chemotaxis}, 
with $\bm{r}_a(t)$ denoting the trajectory of the $a$th cell.  
 {This local continuity equation gives the stochastic version of the KS model as 
\begin{equation}    \label{eq:stoch-KS}
    \partial_t C +\nabla \cdot (\jD + \JKS) = 0. 
\end{equation}}
Here, the diffusive current $\jD$ is given by the stochastic analogue of the Fick's law, namely 
$\jD = - D\nabla C - \sqrt{2DC} \, \bm{\xi}(\bm{r},t)$
where the noise $\bm{\xi}$ now represents a zero mean, unit variance Gaussian white noise field, 
and the KS contribution $\JKS$ results from the chemotactic drift as $\JKS = \nu_1 C \nabla \Phi$.  
Note that averaging the stochastic KS equation with respect to noise realisations recovers the original mean-field KS chemotactic dynamics~\cite{grima-manybody,chavanis2010stochasticKS}.

Specifically focusing on `self-chemotactic'  systems~\cite{hillen2009,chavanis2010stochasticKS}, in which case the chemical field $\Phi$ is created by the signalling cells themselves~\footnote{In some cases, $\Phi$ may alternatively represent the density of the nutrients in the environment, and its spatial gradients then result from the consumption of the nutrients by the cells.}, 
the chemical concentration is governed by the diffusion equation
\begin{align}   \label{eq:Phi-diffusion}
    \partial_t \Phi = D_{\Phi} \nabla^2 \Phi - \kappa^2 \Phi + \alpha \, C.  
\end{align}
Here the cell density $C$ acts as the source term with  $\alpha$ determining the rate of signal production (or consumption) by the cells, $D_\Phi$ is the diffusion coefficient of the chemical molecules, and $\kappa^2$ encodes their degradation rate in the environment.  
Effective interactions among such cells then result from their chemotactic response to the chemical field  generated by other cells in the colony. 
The long-range, Coulomb nature of these interactions becomes more evident in the limit of fast-diffusing signals, where there is a time-scale separation between the dynamics of the signals and the cells. 
This discrepancy is rather prevalent in eukaryotes, with examples such as \textit{D. discoideum}, microglia cells, and neutrophils for which $D_\Phi/D \sim 10^{2}-10^{3}$~\cite{diffusion-difference1,diffusion-difference2}. 
Upon taking the appropriate limit, and defining the density and chemical fluctuations by 
$\rho(\bm{r},t) = C(\bm{r},t) - C_0$ 
and
$\phi(\bm{r},t) = \Phi(\bm{r},t) - \Phi_0$
where $C_0$ and $\Phi_0$ are the respective mean values~\cite{chavanis-kineticchemo,jager1992explosions}, one arrives at 
\begin{align}   \label{eq:phi-poisson}
    -\nabla^2 \phi (\bm{r},t) = \rho (\bm{r},t).  
\end{align}
This Poisson relation determines the quasi-stationary chemical field $\phi$ as a functional of the cell density fluctuations $\rho$. 

The stochastic KS equation can also be 
expanded accordingly in terms of $\rho$ and $\phi$. 
 {In order to do this, we note that 
$\jD = -D\nabla \rho + \sqrt{2DC_0}\sqrt{1+\rho/C_0} \,  \bm{\xi} \approx -D\nabla \rho + \sqrt{2DC_0} \,\bm{\xi}$
and 
$\JKS = \nu_1 (C_0+\rho) \nabla\phi$. 
Substituting these relations into Eq.~\eqref{eq:stoch-KS}, and on making use of the Poisson relation~\eqref{eq:phi-poisson}, we arrive at 
}
\begin{align}
    \left(\partial_t -D\nabla^2 +\sigma \right) \rho &=
    -\mu_1 \nabla\cdot (\rho\nabla\phi) +
    \zeta (\bm{r},t),   
    \label{eq:KS-fieldeqs}
\end{align}
  where $\sigma = -C_0 \nu_1$. 
   {A further extension of this model takes into account processes that can lead to the stochastic activation and inactivation of the cells' chemotactic response, with the extra assumption that cells in their inactive state are abundant in the system.  
  This extension can be used to model the dynamics of cells such fibroblasts, which are activated only in response to, e.g. local inflammation or cancerous activity~\cite{yeo2018}. 
  These additional activation-inactivation processes can be generically represented as  
  $\mathrm{active} \xrightleftharpoons[\lambda']{\,\, \lambda \,\,} \mathrm{inactive}$. 
  Due to the abundance of the inactive cells, in the leading approximation the density of the inactive cells, denoted by $C_i$, remains fixed; in addition, for $C_0\lambda = C_i \lambda'$, the activation-inactivation processes tend to keep the system in a homogeneous state with density $C_0$, resembling a `homeostatic' state of the self-chemotactic colony.  
  In this case, performing a volume expansion on the corresponding master equation shows that  $\sigma$ will be modified as 
  $\sigma = \lambda - C_0 \nu_1$~\cite{mahdisoltani2021chemotaxis}. 
Moreover, the noise correlations are given by 
    $\expval{\zeta(\bm{r},t) \zeta(\bm{r}',t')} = 
    2 \left(\calD_0 - \calD_2\nabla^2\right) \delta^d(\bm{r}-\bm{r}') \delta(t-t'),$ 
%
where $\calD_0 = \lambda C_0$ represents the strength of the nonconserved part of the noise that arises from the nonconserving  activation-inactivation processes, and $\calD_2 = D C_0$ denotes the strength of the conserved part of the noise, which is due to the noisy trajectories of the cells and is already included in the stochastic KS model. } 

\begin{figure*}[t]
   \centering
  \includegraphics[width=.7\textwidth,keepaspectratio]{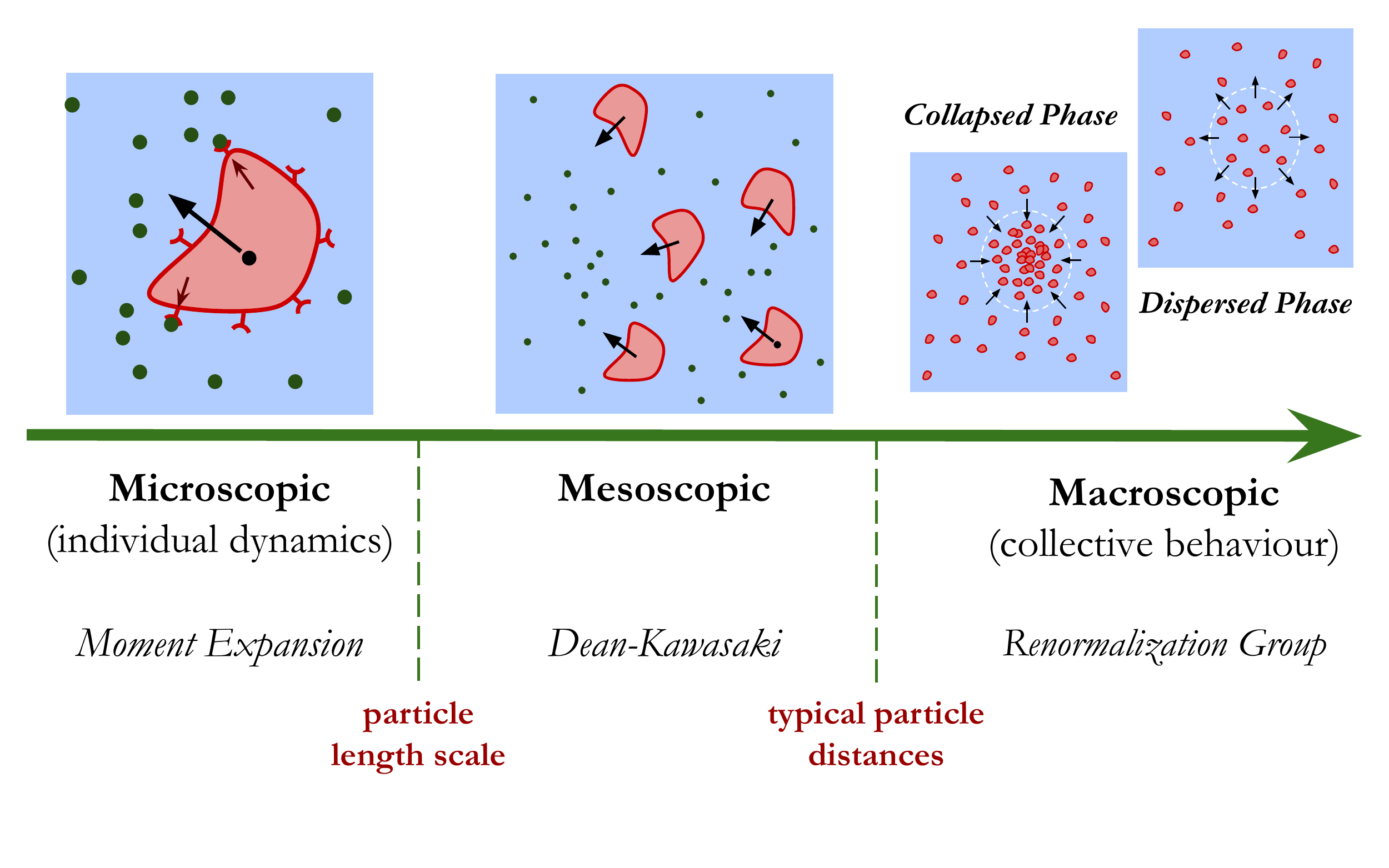}
    \caption{Schematic of a chemotactic system across different scales and the general methods used to investigate each scale. In particular, the polarity-induced mechanism appears as a sub-leading gradient-sensing term at the particle level, but it leads to an RG-relevant term in the macroscopic scales~\cite{mahdisoltani2021chemotaxis}.} 
	\label{fig:chemo-schematic}
\end{figure*}

 {It is worth noting that $\sigma$ in Eq.~\eqref{eq:KS-fieldeqs} determines the rate of the linear growth (or decay) of the density fluctuations in the system, and as such acts as a control parameter akin to, e.g., temperature in an Ising model. 
The control parameter $\sigma$ delimits two regimes (depicted in the macroscopic sector in Fig.~\ref{fig:chemo-schematic}): 
for $\sigma>0$, the chemotactic system is linearly stable with respect to density modulations, and it remains in a `dispersed phase' where density fluctuations around the average density $C_0$ are exponentially suppressed; 
 for $\sigma<0$, density fluctuations are linearly unstable, and the system goes into an inhomogeneous  `collapsed phase' where particles aggregate in high-density clusters. 
 For $\sigma \to 0$, fluctuations become long-lived with a diverging correlation length   $\sqrt{D/\sigma}$; this is the critical regime of the chemotactic dynamics whose asymptotic  large-scale behaviour is dictated by the nonlinear interaction terms. 
 }

Equation~\eqref{eq:KS-fieldeqs} defines the field equations that govern the KS dynamics of a chemotactic system. 
 { To fully capture the macro-scale physics of the chemotactic colony,  it is necessary to analyse the scaling properties of all allowed nonlinear terms in these field equations in addition to the KS nonlinearity.}    
RG-relevant terms that are allowed by physical and symmetry considerations then need to be incorporated in the field equations, since such terms are important in determining the distributions and correlation functions of the system in the asymptotic large-scale limit~\cite{kardar-statfield}.  
 { 
A standard scaling analysis starts with a rescaling of space, time, and density fluctuations according to 
%
    $\bm{r} \to b \bm{r}$, 
    $t \to b^z t$, 
    $\rho \to b^\chi \rho$, 
    and
    $\phi \to b^\psi \phi$, 
%
where $b=e^\ell$ is the scaling factor, $z$ is the so-called dynamic exponent, and $\chi$ and $\psi$ denote the field exponents; note that the Poisson relation~\eqref{eq:phi-poisson} implies the identity $\psi = \chi +2$.     
The mean-field (Gaussian) exponents are then obtained  by requiring the invariance of the linear part of the dynamics, with $\sigma\!=\!0$, under the scale transformation.    
The dynamic exponent in this case takes the diffusive value $z_0 = 2$.  
The Gaussian value of the field exponent $\chi$ depends on how the noise is included in the model:  
in the presence of the nonconserved noise, we get
%
    $\chi_0^{\mathrm{non}} = 1 - \frac{d}{2}$, 
%
rendering the conserved part of the noise  irrelevant (which scales as $b^{-1}$ at the mean-field level).  
On the other hand, if the nonconserved noise is absent (i.e., when the activation-inactivation processes are ignored and $\lambda = 0$), we have
%
    $\chi_0^{\mathrm{con}} = -\frac{d}{2}$.  
}

 {
One can subsequently use the mean-field exponents to determine the relevance of the nonlinear interaction terms in the vicinity of the Gaussian fixed point. 
For the KS nonlinearity  
$\mu_1 \nabla\cdot(\rho\nabla\phi)$, the resulting  scaling factor is readily found as 
$b^{z+\chi}$;  
substituting the exponents with their mean-field values, it can be seen that this term scales as $b^{3-d/2}$ with the nonconserved noise, and therefore it grows under rescaling in 
$d < d_c^{\mathrm{non}} = 6$ 
dimensions, 
while with the conserved noise its scaling is given by $b^{2-d/2}$ and it grows in 
$d<d_c^{\mathrm{con}} = 4$ 
spatial dimensions.  
In both scenarios, the KS nonlinearity is a  relevant interaction for the experimental setups in $d=2$ and $d=3$  dimensions. }

 {
To examine the relevance of other possible nonlinearities, one can consider a general nonlinear term 
%
    $a_{lmn} \nabla^l (\nabla\phi)^m \rho^n$,   
%
which assumes the chemical field $\phi$ enters the chemotactic dynamics only via its gradients.  
For this nonlinear term to be allowed in the Langevin dynamics, it should meet three conditions: 
%
first, $l \geq 1$, since such a term in principle arises from the continuity equation, and therefore it should have at least one gradient operator;   
second, $m+n\geq 2$ with both $m$ and $n$ non-negative, since a nonlinearity should contain at least two fields;  and 
third, $(l+m)$ must be an even number for the interaction term to be a scalar.
}
 { 
The Gaussian scaling dimension of the coupling $a_{lmn}$ added to Eq.~\eqref{eq:KS-fieldeqs} is then computed using the Gaussian exponents $z_0$ and $\chi_0$, and it is straightforward to see that there are only three relevant couplings:} 
(1) the KS interaction term represented by 
$\mu_1 \nabla\cdot(\rho \nabla\phi)$, 
(2) the `polarity-induced' chemotactic term 
$\mu_2 \nabla^2(\nabla\phi)^2$, 
and (3) the `nematic' coupling effect
$\mu_3 \nabla\cdot(\nabla\phi)^3$~\cite{mahdisoltani2021chemotaxis}. 

These nonlinear terms are shown in  Table~\ref{table:relevant-nonlins} along with their symmetry properties that include their gradient structure, free energy functional, and `Galilean' invariance~\cite{mahdisoltani2021chemotaxis}. 
The gradient structure shows the underlying microscopic force field is irrotational; this can be checked using the transformation 
$\nabla \phi \to \nabla\phi + \varepsilon\nabla f(\rho) \times \nabla \phi$ 
performed on the chemical field. A direct calculation then reveals that only the $\mu_3$ term acquires a term linear in the parameter $\varepsilon$ and thus breaks the gradient structure,  while the other two couplings are unaffected by the transformation at the linear order.   
%
%

It is also straightforward to show that, among the relevant terms, only the KS term has a free-energy structure, namely its current can be written as 
$ -\mu_1 \rho \nabla \frac{\delta \calFKS}{\delta\rho}$
with the free energy given by 
%
    $\calF_{\mathrm{KS}} [\rho] = -\frac{1}{2} 
    \int \dif^d \bm{r} \, \rho \, \phi 
    = -\frac{1}{2} \int \dif^d \bm{r} \, \dif^d \bm{r}' \,
    \rho(\bm{r}) \, K_c(\bm{r}-\bm{r}') \, \rho(\bm{r}'),$ 
%
and $K_c(\bm{r})$ representing the Coulomb kernel that satisfies  
$-\nabla^2 K_c (\bm{r}-\bm{r}') = \delta^d(\bm{r}-\bm{r}')$.  
The KS interaction thus has the tendency to minimize the associated free energy by evolving the system toward a  distribution governed by Boltzmann statistics in the long-time limit~\cite{kardar-statfield,tauber}.
The $\mu_2$ and $\mu_3$ relevant couplings, however, do not correspond to an underlying functional, and therefore they tend to evolve the system toward a nonequilibrium steady-state distribution that is in general different from the equilibrium KS system.
%
%

Finally, one may study how the chemotactic interactions behave under a `Galilean' transformation, which is defined  similar to that known from the Kardar--Parisi--Zhange (KPZ)  equation~\cite{kardar1986,medina1989}, namely 
\begin{subequations} \label{eq:galilean-transform}
\begin{align}
    \phi'(\bm{r},t) &= 
    \phi\big(\bm{r}+  t u_{\rm G}  \bm{w} , t\big)-\bm{w}\cdot\bm{r},
        \\
    \rho'(\bm{r},t) &= 
    \rho \big(\bm{r}+ t u_{\rm G} \bm{w} ,t\big), 
\end{align}
\end{subequations}
where $\bm{w}$ is any arbitrary vector and $u_{\rm G}$ is the parameter of the Galilean transformation.  
 {This transformation is essentially a coordinate boost together with the addition of a  constant term to the chemical gradient. 
By applying the Galilean transformation to Eq.~\eqref{eq:KS-fieldeqs} 
through making the changes 
$\partial_t \to \partial_t + u_G \bm{w}\cdot\nabla$, 
$\nabla \to \nabla$,  
$\phi \to \phi - \bm{w}\cdot\bm{r}$, and  
$\rho \to \rho$, 
one can see that in the presence of $\mu_1$ and $\mu_2$ terms only, choosing 
$ u_{\rm G} = \mu_1 - 2\mu_2$ keeps the full dynamics unaffected; therefore, these couplings  represent \textit~{Galilean invariant} nonlinearities.   
The $\mu_3$ coupling, however, is not Galilean invariant, as it would change the structure of the dynamics under the transformation.  
}
%

%
%
%

\begin{table*}[t]
\centering
        \begin{tabular}{P{2.25cm}P{2.25cm}P{2.25cm}P{2.25cm}P{6cm}}
        \hline\hline \\[-.1cm]
          \multirow{2}{2cm}{\textbf{nonlinear coupling}}  & 
           \multirow{2}{2cm}{\textbf{Gradient structure}} &
           \multirow{2}{2cm}{\textbf{Free energy}} &
          \textbf{Galilean invariant} & \multirow{2}{2cm}{\textbf{Comments}}   
          \\[.5cm]
        \hline \\[-.3cm]
        $\mu_1\nabla\!\cdot\!(\rho\nabla\phi)$ &
        Yes &
        Yes &
        Yes &
        \KS chemotaxis 
        \\[.15cm]
         $\mu_2\nabla^2 (\nabla\phi)^2 $  & 
         Yes & 
         No &
         Yes &
         Polarity-induced chemotaxis
         \\[.15cm]
         $\mu_3\nabla\!\cdot\!(\nabla\phi)^3$ &
         No &
         No &
         No &
         Self-propulsion/nematic order
         \\[.1cm]
        \hline\hline
        \end{tabular}
        \caption{Relevant nonlinear couplings at the upper critical dimension 
        ($d^{\rm con}_c=4$ and $d^{\rm non}_c=6$) and their symmetry properties regarding the gradient structure, which shows whether the underlying force field is rotational, the free energy structure, which determines the interaction is equilibrium or nonequilibrium, and the emergent Galilean invariance.}
        \label{table:relevant-nonlins}
\end{table*}  

These symmetries restrict the Langevin theory in specific ways; in particular, limiting the nonlinear couplings to $\mu_1$ and $\mu_2$ only guarantees that  the $\mu_3$ coupling is not generated from coarse graining the theory. The dynamics under consideration is thus 
\begin{align}   \label{eq:Langevin-mu1mu2}
    \left(\partial_t -D\nabla^2 + \sigma \right)\rho = 
    -\mu_1 \nabla\cdot(\rho\nabla\phi) -\mu_2 \nabla^2 (\nabla\phi)^2 + \zeta (\bm{r},t). 
\end{align}
As noted earlier, this dynamics defines a critical point, which is reached by tuning $\sigma = 0$, and it separates the collapsed phase of the system ($\sigma < 0$) from the dispersed phase ($\sigma >0$), see Fig.~\ref{fig:chemo-schematic}. 
At this critical point, the correlation length of the fluctuation modes diverges and the system becomes statistically self similar. 
Consequently, the large-scale density correlation function takes a generic scaling form as~\cite{tauber,medina1989,barabasi} 
\begin{align} \label{eq:rhorhoscalingform}
    \left\langle \rho(\bm{x},t) \rho(\bm{x}',t') \right\rangle \sim |\bm{x}-\bm{x}'|^{2\chi} \, 
    F\left( \frac{|t-t'|}{|\bm{x}-\bm{x}'|^z} \right), 
\end{align}
where $F$ is the scaling function. 
For the Gaussian theory in the absence of  nonlinearities, the mean-field values of the scaling exponents are determined by the requirement that the corresponding Langevin dynamics remain unchanged upon change of scale; including the relevant  nonlinear terms in the theory modifies the values of these exponents, and their new values at the associated RG fixed points must be evaluated through the coarse-graining and rescaling steps of the RG procedure~\cite{kardar-statfield,tauber}.

The details of the RG calculation are rather lengthy and can be found in Ref.~\cite{mahdisoltani2021chemotaxis}. 
The unique symmetry features of the theory, however, directly lead to exact values for the scaling exponents. 
%
First, note that the perturbative corrections to the chemotactic couplings $\mu_{1,2}$ should respect the Galilean invariance of the theory, and therefore the symmetry parameter $u_{\rm G}=\mu_1-2\mu_2$ must remain fixed along the RG flow lines. 
As a result, the corresponding scaling factor should be equal to unit; 
 {since the scaling factor for both $\mu_1$ and $\mu_2$ is given by $b^{z+\chi}$,} the Galilean symmetry then translates to the exact exponent identity 
%
    $z + \chi = 0$~\cite{mahdisoltani2021chemotaxis,tauber}. 
%
Additionally, the structure of the perturbation theory is such that the leading noise corrections are $\propto k^4$, and therefore they are subdominant to the noise in Eq.~\eqref{eq:Langevin-mu1mu2} and do not affect its correlation structure in the macroscopic limit. 
This leads to a second exponent identity, satisfied at RG fixed points, namely 
    $z^{\rm con}-2\chi^{\rm con} = 2+d$ for dynamics with the conserved noise only ($\calD_0 = 0$ and $\calD_2 \neq 0$) and 
     $z^{\rm non}-2\chi^{\rm non} = d$ for dynamics with the non-conserved noise ($\calD_0 \neq 0$)\footnote{In the presence of the non-conserved noise, the conserved part is irrelevant and can be discarded.}.   
%
%
These identities completely determine the fixed point value of the scaling exponents as  
\begin{align}
z^{\rm con} = - \chi^{\rm con} = \frac{d+2}{3}
\qquad \text{and} \qquad
z^{\rm non}=-\chi^{\rm non}= \frac{d}{3},  
\end{align} 
for the conserved and nonconserved noises, respectively. 
These scaling exponents hold in dimensions below the respective upper critical dimensions ($d_c^{\rm con}=4$ and $d_c^{\rm non}=6$). 
In both cases, the exponents deviate considerably from the corresponding mean-field values ($z_0 = 2$, $\chi_0^{\rm con}=-\frac{d}{2}$, $\chi_0^{\rm non} = 1-\frac{d}{2}$).  

The scaling exponents evaluated above have a physical interpretation in terms of how density fluctuations propagate over time, and how number fluctuations in sub-volumes of the system scale with the number of particles in that volume. 
The mean-squared displacement (MSD) of the density fluctuations can be characterised by the exponent $\alpha$ as 
%
$\Delta L^2 \equiv\left\langle \bm{x}(t)^2 \right\rangle \sim t^\alpha,$ 
%
where $\alpha \equiv 2/z$. 
With chemotactic interactions, $\alpha$ is given by
\begin{align} \label{eq:alphaexpon-values}
    \alpha^{\rm con} = \frac{6}{d+2},
    \qquad\quad \text{and} \qquad \quad
    \alpha^{\rm non} = \frac{6}{d} \,\,.  
\end{align}
revealing the density fluctuations exhibit anomalous (super-)diffusion with $\alpha > 1$ below the upper critical dimension. 


In a similar fashion, the scaling of the number fluctuations can be written as 
$\Delta N \sim N^\gamma$
with 
$\gamma \equiv 1+ \chi/d$. 
Substituting for the value of $\chi$, one obtains
\begin{align}   \label{eq:gamma-expon-value}
    \gamma^{\rm con} = \frac{2}{3} \left(1-\frac{1}{d}\right), 
    \qquad \text{and} \qquad 
    \gamma^{\rm non} = \frac{2}{3}. 
\end{align}
Note that for the the Gaussian dynamics with conserved noise, 
$\gamma = 1/2$ represents uncorrelated Poissonian fluctuations ($\Delta N \sim \sqrt{N}$). 
With the chemotactic interactions and conserved noise, 
$\gamma^{\rm con}<\gamma^{\rm con}_0=1/2$, which indicates a \textit{hyperuniform} density distribution. In the presence of non-conserved noise, strong fluctuations are already present at the level of the  Gaussian theory since $\gamma_0^{\rm non} = 1/2 +1/d$. 
With the long-ranged chemotactic interactions, the fluctuations will still be suppressed, as 
$\gamma^{\rm non} =2/3 <\gamma^{\rm non}_0=1/2+1/d$. 
The system exhibits \textit{giant number fluctuations} since the number fluctuations still remain stronger than the Poisson form. 
%

\subsection{Remarks}
%
The nonlinear couplings of Table~\ref{table:relevant-nonlins} extend the macroscopic dynamics of the KS model and modify its large-scale features such as the scaling exponents. 
To gain insights into how such nonlinearities may arise from chemotactic effects at the single-cell level, one can use a heuristic analogy with the multipole expansion in electrostatics.  
It is known that eukaryotic cells perform measurements of chemical concentration through receptor proteins that are spread uniformly across their  surface~\cite{levine2013,servant-dynamics99}.  
In the presence of a nonuniform signal concentration in the environment, the distribution of the bound receptors across the membrane become asymmetric, leading to an effective polarisation of the cell 
through actin polymerization~\cite{devreotes2003eukaryotic,iglesias2008}.
The resulting cell polarity, which can be manifest in, e.g., morphology or sensitivity of the cell, 
can affect how the cell moves in the environment~\cite{iglesias2008}. 
%

A hand-waving extension of the KS response in such cases, based on analogy with multipole expansion,  regards  the KS term as the `monopole' response of the cell, while the `dipole' contribution to cell velocity arising from its polarity is  expected to have the form 
    $\bm{v}_p \propto \bm{n}\cdot\nabla\nabla\phi$, 
which vanishes in a constant-gradient chemical concentration.    
The heuristic can be taken further by computing the  `quadrupole'  contribution to cell velocity as
$\bm{v}_Q \propto \bm{n}\bm{n}\cdot\nabla\phi$.
%
%
For \textit{extensible particles}, one can assume that  initially non-polar cells become elongated quickly when put in a chemical gradient;  
the (time-)average of the cell polarity within a linear approximation is then directly proportional to the chemical gradient, i.e. 
$\expval{\bm{n}}\propto\nabla\phi$. 
For \textit{polar particles}, namely those that are  intrinsically polar or have a persistent induced polarity,  
an extension of the multipole analogy to angular velocity yields 
    $\bm{\omega}\propto \bm{n}\times\nabla\phi$;  
this, in turn, suggests that cell polarity vector tends to align with the chemical gradient and thus once more   
$\expval{\bm{n}}\propto \nabla\phi$.  
Substituting the averaged polarity into the dipole and quadrupole expressions gives the mean-field expressions 
$\bm{v}_p \propto \nabla(\nabla\phi)^2$ 
and
$\bm{v}_Q \propto \nabla\phi (\nabla\phi)^2$ for the dipolar and quadrapolar velocity contributions, respectively. 
Noting that the associated particles currents are obtained from multiplying the velocity field with the  cell density, it is straightforward to recover the $\mu_2$ and $\mu_3$ couplings from the above expressions\footnote{This hand-waving argument can be made more rigorous through a moment expansion approach, as discussed in Ref.~\cite{mahdisoltani2021chemotaxis}.}.

\section{Summary and concluding remarks}  \label{sec:conc}

We reviewed some aspects of strong nonequilibrium fluctuations in two model systems with long-ranged Coulomb interactions:  
driven electrolytes and self-chemotactic colonies.  
In Section~\ref{sec:electro}, it was outlined how the  DK stochastic equation along with scaling analysis provide equations that govern the macroscopic dynamics of density and charge fluctuations in a driven electrolyte. The effective anisotropic dynamics exhibits unscreened, long-ranged correlations in the context of generic scale invariance. 
Such correlations in general lead to Casimir-like FIFs on confining boundaries. 
The resulting nonequilibrium pressure acting on the walls in a flat Casimir geometry become long-ranged with algebraic transient parts. 
The magnitude and direction of the FIF could be tuned by varying the external field, and the force can become repulsive in certain parameter regimes. 

In Section~\ref{sec:chemo}, scaling analysis of the KS chemotactic model pointed to additional chemotactic  couplings in the macroscopic Langevin dynamics stemming from, e.g., the cell polarity, self-propulsion, and nematic effects of the particles. 
The specific symmetry structure of the chemotactic theory with the KS and polarity-induced couplings provide the scaling exponents exactly;
these exponents show that density fluctuations at the critical state of the system are super-diffusive, and the colony has non-Poissonian number fluctuations.

A direct extension of the analysis presented in Section~\ref{sec:electro} would be to study how the  perpendicular component of an alternating electric field affects the nonequilibrium force.  
To generalise the discussion in Section~\ref{sec:chemo}, one needs to take into account the nematic effects and self-propulsion of the particles, which are represented at the macroscopic level by the $\mu_3$ coupling.    
%
%

Lastly, analysis of Section~\ref{sec:chemo} has recently been extended to also account for nonlinear logistic birth and death  processes~\cite{zinati-cg2021}, which is known to involve a nonlinearity that is as relevant as the chemotactic nonlinear terms in the RG sense \cite{gelimson2015}. The logistic growth change the structure of the chemotactic field equations in that the exponent identity due to the non-renormalisation of the noise will no longer hold, and therefore the exponents need to be deduced from the RG fixed-point conditions using perturbative methods. 
At the leading (one-loop) order of the perturbation, however, the RG flows exhibit runaway behaviour towards large value of the growth parameter.  
Further investigation of the chemotactic dynamics in the presence of nonlinear birth and death processes  would be relevant to modelling metastasising cancer cells and will inform on the crucial role of growth and chemotaxis in emergent large-scale properties of dividing cell colonies.

\section{Acknowledgements}
We acknowledge fruitful collaborations on some of the topics presented here with C. Duclut, A. Gambassi, and R.B.A. Zinati. This work has received support from the Max Planck School Matter to Life and the MaxSynBio Consortium, which are jointly funded by the Federal Ministry of Education and Research (BMBF) of Germany, and the Max Planck Society.

\bibliography{v2-reference.bib}

\begin{thebibliography}{92}%
\makeatletter
\providecommand \@ifxundefined [1]{%
 \@ifx{#1\undefined}
}%
\providecommand \@ifnum [1]{%
 \ifnum #1\expandafter \@firstoftwo
 \else \expandafter \@secondoftwo
 \fi
}%
\providecommand \@ifx [1]{%
 \ifx #1\expandafter \@firstoftwo
 \else \expandafter \@secondoftwo
 \fi
}%
\providecommand \natexlab [1]{#1}%
\providecommand \enquote  [1]{``#1''}%
\providecommand \bibnamefont  [1]{#1}%
\providecommand \bibfnamefont [1]{#1}%
\providecommand \citenamefont [1]{#1}%
\providecommand \href@noop [0]{\@secondoftwo}%
\providecommand \href [0]{\begingroup \@sanitize@url \@href}%
\providecommand \@href[1]{\@@startlink{#1}\@@href}%
\providecommand \@@href[1]{\endgroup#1\@@endlink}%
\providecommand \@sanitize@url [0]{\catcode `\\12\catcode `\$12\catcode
  `\&12\catcode `\#12\catcode `\^12\catcode `\_12\catcode `\%12\relax}%
\providecommand \@@startlink[1]{}%
\providecommand \@@endlink[0]{}%
\providecommand \url  [0]{\begingroup\@sanitize@url \@url }%
\providecommand \@url [1]{\endgroup\@href {#1}{\urlprefix }}%
\providecommand \urlprefix  [0]{URL }%
\providecommand \Eprint [0]{\href }%
\providecommand \doibase [0]{https://doi.org/}%
\providecommand \selectlanguage [0]{\@gobble}%
\providecommand \bibinfo  [0]{\@secondoftwo}%
\providecommand \bibfield  [0]{\@secondoftwo}%
\providecommand \translation [1]{[#1]}%
\providecommand \BibitemOpen [0]{}%
\providecommand \bibitemStop [0]{}%
\providecommand \bibitemNoStop [0]{.\EOS\space}%
\providecommand \EOS [0]{\spacefactor3000\relax}%
\providecommand \BibitemShut  [1]{\csname bibitem#1\endcsname}%
\let\auto@bib@innerbib\@empty
\bibitem [{\citenamefont {Gupta}\ and\ \citenamefont
  {Ruffo}(2017)}]{LRI-birdview}%
  \BibitemOpen
  \bibfield  {author} {\bibinfo {author} {\bibfnamefont {S.}~\bibnamefont
  {Gupta}}\ and\ \bibinfo {author} {\bibfnamefont {S.}~\bibnamefont {Ruffo}},\
  }\bibfield  {title} {\bibinfo {title} {The world of long-range interactions:
  A bird’s eye view},\ }\href
  {https://www.worldscientific.com/doi/abs/10.1142/S0217751X17410184}
  {\bibfield  {journal} {\bibinfo  {journal} {Int. J. Mod. Phys. A}\ }\textbf
  {\bibinfo {volume} {32}},\ \bibinfo {pages} {1741018} (\bibinfo {year}
  {2017})}\BibitemShut {NoStop}%
\bibitem [{\citenamefont {Chavanis}(2008{\natexlab{a}})}]{chavanis-LRI-V}%
  \BibitemOpen
  \bibfield  {author} {\bibinfo {author} {\bibfnamefont {P.-H.}\ \bibnamefont
  {Chavanis}},\ }\bibfield  {title} {\bibinfo {title} {Hamiltonian and brownian
  systems with long-range interactions: V. stochastic kinetic equations and
  theory of fluctuations},\ }\href
  {https://www.sciencedirect.com/science/article/abs/pii/S037843710800527X}
  {\bibfield  {journal} {\bibinfo  {journal} {Physica A}\ }\textbf {\bibinfo
  {volume} {387}},\ \bibinfo {pages} {5716} (\bibinfo {year}
  {2008}{\natexlab{a}})}\BibitemShut {NoStop}%
\bibitem [{\citenamefont {Levin}(2002)}]{levin2002correlations}%
  \BibitemOpen
  \bibfield  {author} {\bibinfo {author} {\bibfnamefont {Y.}~\bibnamefont
  {Levin}},\ }\bibfield  {title} {\bibinfo {title} {Electrostatic correlations:
  from plasma to biology},\ }\href
  {https://iopscience.iop.org/article/10.1088/0034-4885/65/11/201/meta}
  {\bibfield  {journal} {\bibinfo  {journal} {Rep. Prog. Phys.}\ }\textbf
  {\bibinfo {volume} {65}},\ \bibinfo {pages} {1577} (\bibinfo {year}
  {2002})}\BibitemShut {NoStop}%
\bibitem [{\citenamefont {Wright}(2007)}]{wrightelectrolyte}%
  \BibitemOpen
  \bibfield  {author} {\bibinfo {author} {\bibfnamefont {M.~R.}\ \bibnamefont
  {Wright}},\ }\href@noop {} {\emph {\bibinfo {title} {An introduction to
  aqueous electrolyte solutions}}}\ (\bibinfo  {publisher} {John Wiley \&
  Sons},\ \bibinfo {year} {2007})\BibitemShut {NoStop}%
\bibitem [{\citenamefont {Arrhenius}(1887)}]{arrhenius}%
  \BibitemOpen
  \bibfield  {author} {\bibinfo {author} {\bibfnamefont {S.}~\bibnamefont
  {Arrhenius}},\ }\bibfield  {title} {\bibinfo {title} {On the dissociation of
  substances dissolved in water},\ }\href@noop {} {\bibfield  {journal}
  {\bibinfo  {journal} {Z. Phys. Chem}\ }\textbf {\bibinfo {volume} {1}},\
  \bibinfo {pages} {631} (\bibinfo {year} {1887})}\BibitemShut {NoStop}%
\bibitem [{\citenamefont {Debye}\ and\ \citenamefont
  {H\"{u}ckel}(1923)}]{debye1923}%
  \BibitemOpen
  \bibfield  {author} {\bibinfo {author} {\bibfnamefont {P.}~\bibnamefont
  {Debye}}\ and\ \bibinfo {author} {\bibfnamefont {E.}~\bibnamefont
  {H\"{u}ckel}},\ }\bibfield  {title} {\bibinfo {title} {The theory of
  electrolytes i. the lowering of the freezing point and related occurrences},\
  }\href@noop {} {\bibfield  {journal} {\bibinfo  {journal} {Physikalische
  Zeitschrift}\ }\textbf {\bibinfo {volume} {24}},\ \bibinfo {pages} {185}
  (\bibinfo {year} {1923})}\BibitemShut {NoStop}%
\bibitem [{\citenamefont {Kosterlitz}\ and\ \citenamefont
  {Thouless}(1973)}]{kosterlitz}%
  \BibitemOpen
  \bibfield  {author} {\bibinfo {author} {\bibfnamefont {J.~M.}\ \bibnamefont
  {Kosterlitz}}\ and\ \bibinfo {author} {\bibfnamefont {D.~J.}\ \bibnamefont
  {Thouless}},\ }\bibfield  {title} {\bibinfo {title} {Ordering, metastability
  and phase transitions in two-dimensional systems},\ }\href
  {https://iopscience.iop.org/article/10.1088/0022-3719/6/7/010/meta?casa_token=1X7FEXVN-xAAAAAA:jDbzZaNAsXenZE9cFcHJl6Gi_sFFSchxvnb8CA0mRpYwOvSxRCMAsUIdR8V0JA3Jm3Z4GbQBqA}
  {\bibfield  {journal} {\bibinfo  {journal} {J. Phys. Part C Solid}\ }\textbf
  {\bibinfo {volume} {6}},\ \bibinfo {pages} {1181} (\bibinfo {year}
  {1973})}\BibitemShut {NoStop}%
\bibitem [{\citenamefont {Kosterlitz}(2016)}]{kosterlitz-review}%
  \BibitemOpen
  \bibfield  {author} {\bibinfo {author} {\bibfnamefont {J.~M.}\ \bibnamefont
  {Kosterlitz}},\ }\bibfield  {title} {\bibinfo {title} {{Kosterlitz--Thouless
  physics: a review of key issues}},\ }\href
  {https://iopscience.iop.org/article/10.1088/0034-4885/79/2/026001/meta}
  {\bibfield  {journal} {\bibinfo  {journal} {Rep. Prog. Phys.}\ }\textbf
  {\bibinfo {volume} {79}},\ \bibinfo {pages} {026001} (\bibinfo {year}
  {2016})}\BibitemShut {NoStop}%
\bibitem [{\citenamefont {Marchetti}\ \emph {et~al.}(2013)\citenamefont
  {Marchetti}, \citenamefont {Joanny}, \citenamefont {Ramaswamy}, \citenamefont
  {Liverpool}, \citenamefont {Prost}, \citenamefont {Rao},\ and\ \citenamefont
  {Simha}}]{marchetti2013}%
  \BibitemOpen
  \bibfield  {author} {\bibinfo {author} {\bibfnamefont {M.~C.}\ \bibnamefont
  {Marchetti}}, \bibinfo {author} {\bibfnamefont {J.~F.}\ \bibnamefont
  {Joanny}}, \bibinfo {author} {\bibfnamefont {S.}~\bibnamefont {Ramaswamy}},
  \bibinfo {author} {\bibfnamefont {T.~B.}\ \bibnamefont {Liverpool}}, \bibinfo
  {author} {\bibfnamefont {J.}~\bibnamefont {Prost}}, \bibinfo {author}
  {\bibfnamefont {M.}~\bibnamefont {Rao}},\ and\ \bibinfo {author}
  {\bibfnamefont {R.~A.}\ \bibnamefont {Simha}},\ }\bibfield  {title} {\bibinfo
  {title} {Hydrodynamics of soft active matter},\ }\href
  {https://doi.org/10.1103/RevModPhys.85.1143} {\bibfield  {journal} {\bibinfo
  {journal} {Rev. Mod. Phys.}\ }\textbf {\bibinfo {volume} {85}},\ \bibinfo
  {pages} {1143} (\bibinfo {year} {2013})}\BibitemShut {NoStop}%
\bibitem [{\citenamefont {Ramaswamy}(2010)}]{ramaswamy-rev}%
  \BibitemOpen
  \bibfield  {author} {\bibinfo {author} {\bibfnamefont {S.}~\bibnamefont
  {Ramaswamy}},\ }\bibfield  {title} {\bibinfo {title} {The mechanics and
  statistics of active matter},\ }\href
  {https://www.annualreviews.org/doi/abs/10.1146/annurev-conmatphys-070909-104101}
  {\bibfield  {journal} {\bibinfo  {journal} {Annu. Rev. Condens. Matter
  Phys.}\ }\textbf {\bibinfo {volume} {1}},\ \bibinfo {pages} {323} (\bibinfo
  {year} {2010})}\BibitemShut {NoStop}%
\bibitem [{\citenamefont {Barton}\ and\ \citenamefont
  {Barton}(1989)}]{barton1989}%
  \BibitemOpen
  \bibfield  {author} {\bibinfo {author} {\bibfnamefont {G.}~\bibnamefont
  {Barton}}\ and\ \bibinfo {author} {\bibfnamefont {G.}~\bibnamefont
  {Barton}},\ }\href@noop {} {\emph {\bibinfo {title} {Elements of Green's
  functions and propagation: potentials, diffusion, and waves}}}\ (\bibinfo
  {publisher} {Oxford University Press},\ \bibinfo {year} {1989})\BibitemShut
  {NoStop}%
\bibitem [{\citenamefont {Golestanian}(2012)}]{golestanian2012}%
  \BibitemOpen
  \bibfield  {author} {\bibinfo {author} {\bibfnamefont {R.}~\bibnamefont
  {Golestanian}},\ }\bibfield  {title} {\bibinfo {title} {Collective
  {{Behavior}} of {{Thermally Active Colloids}}},\ }\href
  {https://doi.org/10.1103/PhysRevLett.108.038303} {\bibfield  {journal}
  {\bibinfo  {journal} {Phys. Rev. Lett.}\ }\textbf {\bibinfo {volume} {108}},\
  \bibinfo {pages} {038303} (\bibinfo {year} {2012})}\BibitemShut {NoStop}%
\bibitem [{\citenamefont {Tsori}\ and\ \citenamefont
  {de~Gennes}(2004)}]{tsori2004}%
  \BibitemOpen
  \bibfield  {author} {\bibinfo {author} {\bibfnamefont {Y.}~\bibnamefont
  {Tsori}}\ and\ \bibinfo {author} {\bibfnamefont {P.-G.}\ \bibnamefont
  {de~Gennes}},\ }\bibfield  {title} {\bibinfo {title} {Self-trapping of a
  single bacterium in its own chemoattractant},\ }\href
  {https://doi.org/10.1209/epl/i2003-10237-5} {\bibfield  {journal} {\bibinfo
  {journal} {Europhys. Lett.}\ }\textbf {\bibinfo {volume} {66}},\ \bibinfo
  {pages} {599} (\bibinfo {year} {2004})}\BibitemShut {NoStop}%
\bibitem [{\citenamefont {Chavanis}(2008{\natexlab{b}})}]{chavanis2008}%
  \BibitemOpen
  \bibfield  {author} {\bibinfo {author} {\bibfnamefont {P.-H.}\ \bibnamefont
  {Chavanis}},\ }\bibfield  {title} {\bibinfo {title} {Nonlinear mean field
  {{Fokker}}-{{Planck}} equations. {{Application}} to the chemotaxis of
  biological populations},\ }\href {https://doi.org/10.1140/epjb/e2008-00142-9}
  {\bibfield  {journal} {\bibinfo  {journal} {Eur. Phys. J. B}\ }\textbf
  {\bibinfo {volume} {62}},\ \bibinfo {pages} {179} (\bibinfo {year}
  {2008}{\natexlab{b}})}\BibitemShut {NoStop}%
\bibitem [{\citenamefont {Golestanian}(2009)}]{golestanian2009anomalous}%
  \BibitemOpen
  \bibfield  {author} {\bibinfo {author} {\bibfnamefont {R.}~\bibnamefont
  {Golestanian}},\ }\bibfield  {title} {\bibinfo {title} {Anomalous diffusion
  of symmetric and asymmetric active colloids},\ }\href
  {https://journals.aps.org/prl/abstract/10.1103/PhysRevLett.102.188305}
  {\bibfield  {journal} {\bibinfo  {journal} {Phys. Rev. Lett.}\ }\textbf
  {\bibinfo {volume} {102}},\ \bibinfo {pages} {188305} (\bibinfo {year}
  {2009})}\BibitemShut {NoStop}%
\bibitem [{\citenamefont {Chavanis}\ and\ \citenamefont
  {Sire}(2008)}]{chavanis-jeans-2008}%
  \BibitemOpen
  \bibfield  {author} {\bibinfo {author} {\bibfnamefont {P.-H.}\ \bibnamefont
  {Chavanis}}\ and\ \bibinfo {author} {\bibfnamefont {C.}~\bibnamefont
  {Sire}},\ }\bibfield  {title} {\bibinfo {title} {Jeans type analysis of
  chemotactic collapse},\ }\href
  {https://www.sciencedirect.com/science/article/abs/pii/S0378437108001453}
  {\bibfield  {journal} {\bibinfo  {journal} {Physica A}\ }\textbf {\bibinfo
  {volume} {387}},\ \bibinfo {pages} {4033} (\bibinfo {year}
  {2008})}\BibitemShut {NoStop}%
\bibitem [{\citenamefont {Gelimson}\ and\ \citenamefont
  {Golestanian}(2015)}]{gelimson2015}%
  \BibitemOpen
  \bibfield  {author} {\bibinfo {author} {\bibfnamefont {A.}~\bibnamefont
  {Gelimson}}\ and\ \bibinfo {author} {\bibfnamefont {R.}~\bibnamefont
  {Golestanian}},\ }\bibfield  {title} {\bibinfo {title} {Collective
  {{Dynamics}} of {{Dividing Chemotactic Cells}}},\ }\href
  {https://doi.org/10.1103/PhysRevLett.114.028101} {\bibfield  {journal}
  {\bibinfo  {journal} {Phys. Rev. Lett.}\ }\textbf {\bibinfo {volume} {114}},\
  \bibinfo {pages} {028101} (\bibinfo {year} {2015})}\BibitemShut {NoStop}%
\bibitem [{\citenamefont {Agudo-Canalejo}\ and\ \citenamefont
  {Golestanian}(2019)}]{agudo-activephase2019}%
  \BibitemOpen
  \bibfield  {author} {\bibinfo {author} {\bibfnamefont {J.}~\bibnamefont
  {Agudo-Canalejo}}\ and\ \bibinfo {author} {\bibfnamefont {R.}~\bibnamefont
  {Golestanian}},\ }\bibfield  {title} {\bibinfo {title} {Active phase
  separation in mixtures of chemically interacting particles},\ }\href
  {https://journals.aps.org/prl/abstract/10.1103/PhysRevLett.123.018101}
  {\bibfield  {journal} {\bibinfo  {journal} {Phys. Rev. Lett.}\ }\textbf
  {\bibinfo {volume} {123}},\ \bibinfo {pages} {018101} (\bibinfo {year}
  {2019})}\BibitemShut {NoStop}%
\bibitem [{\citenamefont {Golestanian}(2019)}]{golestanian2019}%
  \BibitemOpen
  \bibfield  {author} {\bibinfo {author} {\bibfnamefont {R.}~\bibnamefont
  {Golestanian}},\ }\bibfield  {title} {\bibinfo {title} {Phoretic {{Active
  Matter}}},\ }\href {https://arxiv.org/abs/1909.03747} {\bibfield  {journal}
  {\bibinfo  {journal} {arXiv preprint arXiv:1909.03747}\ } (\bibinfo {year}
  {2019})}\BibitemShut {NoStop}%
\bibitem [{\citenamefont {Goldenfeld}(1992)}]{goldenfeld1992}%
  \BibitemOpen
  \bibfield  {author} {\bibinfo {author} {\bibfnamefont {N.}~\bibnamefont
  {Goldenfeld}},\ }\href@noop {} {\emph {\bibinfo {title} {Lectures on Phase
  Transitions and the Renormalization Group}}}\ (\bibinfo  {publisher}
  {{Perseus Books}},\ \bibinfo {address} {{Reading, Massachusetts}},\ \bibinfo
  {year} {1992})\BibitemShut {NoStop}%
\bibitem [{\citenamefont {Fisher}(1998)}]{fisher-RG-rev}%
  \BibitemOpen
  \bibfield  {author} {\bibinfo {author} {\bibfnamefont {M.~E.}\ \bibnamefont
  {Fisher}},\ }\bibfield  {title} {\bibinfo {title} {Renormalization group
  theory: Its basis and formulation in statistical physics},\ }\href
  {https://journals.aps.org/rmp/abstract/10.1103/RevModPhys.70.653} {\bibfield
  {journal} {\bibinfo  {journal} {Rev. Mod. Phys.}\ }\textbf {\bibinfo {volume}
  {70}},\ \bibinfo {pages} {653} (\bibinfo {year} {1998})}\BibitemShut
  {NoStop}%
\bibitem [{\citenamefont {Stanley}(1999)}]{stanley-RG-rev}%
  \BibitemOpen
  \bibfield  {author} {\bibinfo {author} {\bibfnamefont {H.~E.}\ \bibnamefont
  {Stanley}},\ }\bibfield  {title} {\bibinfo {title} {Scaling, universality,
  and renormalization: Three pillars of modern critical phenomena},\
  }\href@noop {} {\bibfield  {journal} {\bibinfo  {journal} {Rev. Mod. Phys.}\
  }\textbf {\bibinfo {volume} {71}},\ \bibinfo {pages} {S358} (\bibinfo {year}
  {1999})}\BibitemShut {NoStop}%
\bibitem [{\citenamefont {Grinstein}(1991)}]{grinstein91generic}%
  \BibitemOpen
  \bibfield  {author} {\bibinfo {author} {\bibfnamefont {G.}~\bibnamefont
  {Grinstein}},\ }\bibfield  {title} {\bibinfo {title} {Generic scale
  invariance in classical nonequilibrium systems},\ }\href
  {https://aip.scitation.org/doi/10.1063/1.348003} {\bibfield  {journal}
  {\bibinfo  {journal} {J. Appl. Phys.}\ }\textbf {\bibinfo {volume} {69}},\
  \bibinfo {pages} {5441} (\bibinfo {year} {1991})}\BibitemShut {NoStop}%
\bibitem [{\citenamefont {Schmittmann}\ and\ \citenamefont
  {Zia}(1995)}]{zia-schmittmann}%
  \BibitemOpen
  \bibfield  {author} {\bibinfo {author} {\bibfnamefont {B.}~\bibnamefont
  {Schmittmann}}\ and\ \bibinfo {author} {\bibfnamefont {R.}~\bibnamefont
  {Zia}},\ }\bibfield  {title} {\bibinfo {title} {Statistical mechanics of
  driven diffusive systems},\ }in\ \href@noop {} {\emph {\bibinfo {booktitle}
  {Statistical Mechanics of Driven Diffusive System}}},\ \bibinfo {series}
  {Phase Transitions and Critical Phenomena}, Vol.~\bibinfo {volume} {17},\
  \bibinfo {editor} {edited by\ \bibinfo {editor} {\bibfnamefont
  {B.}~\bibnamefont {Schmittmann}}\ and\ \bibinfo {editor} {\bibfnamefont
  {R.}~\bibnamefont {Zia}}}\ (\bibinfo  {publisher} {Academic Press},\ \bibinfo
  {year} {1995})\ pp.\ \bibinfo {pages} {3--214}\BibitemShut {NoStop}%
\bibitem [{\citenamefont {Vicsek}\ and\ \citenamefont
  {Zafeiris}(2012)}]{vicsek-collective-review}%
  \BibitemOpen
  \bibfield  {author} {\bibinfo {author} {\bibfnamefont {T.}~\bibnamefont
  {Vicsek}}\ and\ \bibinfo {author} {\bibfnamefont {A.}~\bibnamefont
  {Zafeiris}},\ }\bibfield  {title} {\bibinfo {title} {Collective motion},\
  }\href
  {https://www.sciencedirect.com/science/article/abs/pii/S0370157312000968}
  {\bibfield  {journal} {\bibinfo  {journal} {Phys. Rep.}\ }\textbf {\bibinfo
  {volume} {517}},\ \bibinfo {pages} {71} (\bibinfo {year} {2012})}\BibitemShut
  {NoStop}%
\bibitem [{\citenamefont {Chialvo}(2010)}]{chialvo}%
  \BibitemOpen
  \bibfield  {author} {\bibinfo {author} {\bibfnamefont {D.~R.}\ \bibnamefont
  {Chialvo}},\ }\bibfield  {title} {\bibinfo {title} {Emergent complex neural
  dynamics},\ }\href {https://www.nature.com/articles/nphys1803} {\bibfield
  {journal} {\bibinfo  {journal} {Nat. Phys.}\ }\textbf {\bibinfo {volume}
  {6}},\ \bibinfo {pages} {744} (\bibinfo {year} {2010})}\BibitemShut {NoStop}%
\bibitem [{\citenamefont {Sornette}(2006)}]{sornette-book}%
  \BibitemOpen
  \bibfield  {author} {\bibinfo {author} {\bibfnamefont {D.}~\bibnamefont
  {Sornette}},\ }\href@noop {} {\emph {\bibinfo {title} {Critical phenomena in
  natural sciences: chaos, fractals, selforganization and disorder: concepts
  and tools}}}\ (\bibinfo  {publisher} {Springer Science \& Business Media},\
  \bibinfo {year} {2006})\BibitemShut {NoStop}%
\bibitem [{\citenamefont {Bak}\ \emph {et~al.}(1988)\citenamefont {Bak},
  \citenamefont {Tang},\ and\ \citenamefont {Wiesenfeld}}]{BTW}%
  \BibitemOpen
  \bibfield  {author} {\bibinfo {author} {\bibfnamefont {P.}~\bibnamefont
  {Bak}}, \bibinfo {author} {\bibfnamefont {C.}~\bibnamefont {Tang}},\ and\
  \bibinfo {author} {\bibfnamefont {K.}~\bibnamefont {Wiesenfeld}},\ }\bibfield
   {title} {\bibinfo {title} {Self-organized criticality},\ }\href
  {https://journals.aps.org/pra/abstract/10.1103/PhysRevA.38.364} {\bibfield
  {journal} {\bibinfo  {journal} {Phys. Rev. A}\ }\textbf {\bibinfo {volume}
  {38}},\ \bibinfo {pages} {364} (\bibinfo {year} {1988})}\BibitemShut
  {NoStop}%
\bibitem [{\citenamefont {R{\'a}cz}(2002)}]{racz-lecturenotes}%
  \BibitemOpen
  \bibfield  {author} {\bibinfo {author} {\bibfnamefont {Z.}~\bibnamefont
  {R{\'a}cz}},\ }\bibfield  {title} {\bibinfo {title} {Nonequilibrium phase
  transitions},\ }\href {https://arxiv.org/abs/cond-mat/0210435} {\bibfield
  {journal} {\bibinfo  {journal} {arXiv preprint cond-mat/0210435}\ } (\bibinfo
  {year} {2002})}\BibitemShut {NoStop}%
\bibitem [{\citenamefont {Kavokine}\ \emph {et~al.}(2020)\citenamefont
  {Kavokine}, \citenamefont {Netz},\ and\ \citenamefont
  {Bocquet}}]{nanofluidsreview}%
  \BibitemOpen
  \bibfield  {author} {\bibinfo {author} {\bibfnamefont {N.}~\bibnamefont
  {Kavokine}}, \bibinfo {author} {\bibfnamefont {R.~R.}\ \bibnamefont {Netz}},\
  and\ \bibinfo {author} {\bibfnamefont {L.}~\bibnamefont {Bocquet}},\
  }\bibfield  {title} {\bibinfo {title} {Fluids at the nanoscale: From
  continuum to subcontinuum transport},\ }\href
  {https://www.annualreviews.org/doi/abs/10.1146/annurev-fluid-071320-095958}
  {\bibfield  {journal} {\bibinfo  {journal} {Annu. Rev. Fluid Mech.}\ }\textbf
  {\bibinfo {volume} {53}} (\bibinfo {year} {2020})}\BibitemShut {NoStop}%
\bibitem [{\citenamefont {Mahdisoltani}\ and\ \citenamefont
  {Golestanian}(2021{\natexlab{a}})}]{mahdisoltani2021long}%
  \BibitemOpen
  \bibfield  {author} {\bibinfo {author} {\bibfnamefont {S.}~\bibnamefont
  {Mahdisoltani}}\ and\ \bibinfo {author} {\bibfnamefont {R.}~\bibnamefont
  {Golestanian}},\ }\bibfield  {title} {\bibinfo {title} {Long-range
  fluctuation-induced forces in driven electrolytes},\ }\href
  {https://journals.aps.org/prl/abstract/10.1103/PhysRevLett.126.158002}
  {\bibfield  {journal} {\bibinfo  {journal} {Phys. Rev. Lett.}\ }\textbf
  {\bibinfo {volume} {126}},\ \bibinfo {pages} {158002} (\bibinfo {year}
  {2021}{\natexlab{a}})}\BibitemShut {NoStop}%
\bibitem [{\citenamefont {Mahdisoltani}\ and\ \citenamefont
  {Golestanian}(2021{\natexlab{b}})}]{mahdisoltani-NJP}%
  \BibitemOpen
  \bibfield  {author} {\bibinfo {author} {\bibfnamefont {S.}~\bibnamefont
  {Mahdisoltani}}\ and\ \bibinfo {author} {\bibfnamefont {R.}~\bibnamefont
  {Golestanian}},\ }\bibfield  {title} {\bibinfo {title} {Transient
  fluctuation-induced forces in driven electrolytes after an electric field
  quench},\ }\href {https://doi.org/10.1088/1367-2630/ac0f1a} {\bibfield
  {journal} {\bibinfo  {journal} {New J. Phys.}\ }\textbf {\bibinfo {volume}
  {23}},\ \bibinfo {pages} {073034} (\bibinfo {year}
  {2021}{\natexlab{b}})}\BibitemShut {NoStop}%
\bibitem [{\citenamefont {Mahdisoltani}\ \emph {et~al.}(2021)\citenamefont
  {Mahdisoltani}, \citenamefont {Zinati}, \citenamefont {Duclut}, \citenamefont
  {Gambassi},\ and\ \citenamefont {Golestanian}}]{mahdisoltani2021chemotaxis}%
  \BibitemOpen
  \bibfield  {author} {\bibinfo {author} {\bibfnamefont {S.}~\bibnamefont
  {Mahdisoltani}}, \bibinfo {author} {\bibfnamefont {R.~B.~A.}\ \bibnamefont
  {Zinati}}, \bibinfo {author} {\bibfnamefont {C.}~\bibnamefont {Duclut}},
  \bibinfo {author} {\bibfnamefont {A.}~\bibnamefont {Gambassi}},\ and\
  \bibinfo {author} {\bibfnamefont {R.}~\bibnamefont {Golestanian}},\
  }\bibfield  {title} {\bibinfo {title} {Nonequilibrium polarity-induced
  chemotaxis: Emergent galilean symmetry and exact scaling exponents},\ }\href
  {https://journals.aps.org/prresearch/abstract/10.1103/PhysRevResearch.3.013100}
  {\bibfield  {journal} {\bibinfo  {journal} {Phys. Rev. Research}\ }\textbf
  {\bibinfo {volume} {3}},\ \bibinfo {pages} {013100} (\bibinfo {year}
  {2021})}\BibitemShut {NoStop}%
\bibitem [{\citenamefont {Zinati}\ \emph {et~al.}(2022)\citenamefont {Zinati},
  \citenamefont {Duclut}, \citenamefont {Mahdisoltani}, \citenamefont
  {Gambassi},\ and\ \citenamefont {Golestanian}}]{zinati-cg2021}%
  \BibitemOpen
  \bibfield  {author} {\bibinfo {author} {\bibfnamefont {R.~B.~A.}\
  \bibnamefont {Zinati}}, \bibinfo {author} {\bibfnamefont {C.}~\bibnamefont
  {Duclut}}, \bibinfo {author} {\bibfnamefont {S.}~\bibnamefont
  {Mahdisoltani}}, \bibinfo {author} {\bibfnamefont {A.}~\bibnamefont
  {Gambassi}},\ and\ \bibinfo {author} {\bibfnamefont {R.}~\bibnamefont
  {Golestanian}},\ }\bibfield  {title} {\bibinfo {title} {Stochastic dynamics
  of chemotactic colonies with logistic growth},\ }\href
  {https://doi.org/10.1209/0295-5075/ac48c9} {\bibfield  {journal} {\bibinfo
  {journal} {Europhys. Lett.}\ }\textbf {\bibinfo {volume} {136}},\ \bibinfo
  {pages} {50003} (\bibinfo {year} {2022})}\BibitemShut {NoStop}%
\bibitem [{\citenamefont {T{\"a}uber}(2014)}]{tauber}%
  \BibitemOpen
  \bibfield  {author} {\bibinfo {author} {\bibfnamefont {U.~C.}\ \bibnamefont
  {T{\"a}uber}},\ }\href@noop {} {\emph {\bibinfo {title} {Critical dynamics: a
  field theory approach to equilibrium and non-equilibrium scaling behavior}}}\
  (\bibinfo  {publisher} {Cambridge University Press},\ \bibinfo {year}
  {2014})\BibitemShut {NoStop}%
\bibitem [{\citenamefont {Alder}\ and\ \citenamefont
  {Wainwright}(1970)}]{R1-1}%
  \BibitemOpen
  \bibfield  {author} {\bibinfo {author} {\bibfnamefont {B.~J.}\ \bibnamefont
  {Alder}}\ and\ \bibinfo {author} {\bibfnamefont {T.~E.}\ \bibnamefont
  {Wainwright}},\ }\bibfield  {title} {\bibinfo {title} {Decay of the velocity
  autocorrelation function},\ }\href {https://doi.org/10.1103/PhysRevA.1.18}
  {\bibfield  {journal} {\bibinfo  {journal} {Phys. Rev. A}\ }\textbf {\bibinfo
  {volume} {1}},\ \bibinfo {pages} {18} (\bibinfo {year} {1970})}\BibitemShut
  {NoStop}%
\bibitem [{\citenamefont {Levin}\ \emph {et~al.}(1993)\citenamefont {Levin},
  \citenamefont {Mundy},\ and\ \citenamefont {Dawson}}]{R1-2}%
  \BibitemOpen
  \bibfield  {author} {\bibinfo {author} {\bibfnamefont {Y.}~\bibnamefont
  {Levin}}, \bibinfo {author} {\bibfnamefont {C.}~\bibnamefont {Mundy}},\ and\
  \bibinfo {author} {\bibfnamefont {K.}~\bibnamefont {Dawson}},\ }\bibfield
  {title} {\bibinfo {title} {Relaxation phenomena in self-assembled systems},\
  }\href {https://doi.org/https://doi.org/10.1016/0378-4371(93)90599-Y}
  {\bibfield  {journal} {\bibinfo  {journal} {Physica A}\ }\textbf {\bibinfo
  {volume} {196}},\ \bibinfo {pages} {173} (\bibinfo {year}
  {1993})}\BibitemShut {NoStop}%
\bibitem [{\citenamefont {Perez-Martinez}\ and\ \citenamefont
  {Perkin}(2019)}]{perez2019surface}%
  \BibitemOpen
  \bibfield  {author} {\bibinfo {author} {\bibfnamefont {C.~S.}\ \bibnamefont
  {Perez-Martinez}}\ and\ \bibinfo {author} {\bibfnamefont {S.}~\bibnamefont
  {Perkin}},\ }\bibfield  {title} {\bibinfo {title} {Surface forces generated
  by the action of electric fields across liquid films},\ }\href
  {https://pubs.rsc.org/en/content/articlelanding/2019/sm/c9sm00143c#!divAbstract}
  {\bibfield  {journal} {\bibinfo  {journal} {Soft Matter}\ }\textbf {\bibinfo
  {volume} {15}},\ \bibinfo {pages} {4255} (\bibinfo {year}
  {2019})}\BibitemShut {NoStop}%
\bibitem [{\citenamefont {Long}\ and\ \citenamefont
  {Ajdari}(2001)}]{ajdarinote}%
  \BibitemOpen
  \bibfield  {author} {\bibinfo {author} {\bibfnamefont {D.}~\bibnamefont
  {Long}}\ and\ \bibinfo {author} {\bibfnamefont {A.}~\bibnamefont {Ajdari}},\
  }\bibfield  {title} {\bibinfo {title} {A note on the screening of
  hydrodynamic interactions, in electrophoresis, and in porous media},\ }\href
  {https://link.springer.com/article/10.1007\%2Fs101890170139#citeas}
  {\bibfield  {journal} {\bibinfo  {journal} {Eur. Phys. J. E}\ }\textbf
  {\bibinfo {volume} {4}},\ \bibinfo {pages} {29} (\bibinfo {year}
  {2001})}\BibitemShut {NoStop}%
\bibitem [{\citenamefont {Onsager}\ and\ \citenamefont
  {Fuoss}(1932)}]{onsagerlong1932}%
  \BibitemOpen
  \bibfield  {author} {\bibinfo {author} {\bibfnamefont {L.}~\bibnamefont
  {Onsager}}\ and\ \bibinfo {author} {\bibfnamefont {R.~M.}\ \bibnamefont
  {Fuoss}},\ }\bibfield  {title} {\bibinfo {title} {Irreversible processes in
  electrolytes. diffusion, conductance and viscous flow in arbitrary mixtures
  of strong electrolytes},\ }\href {https://doi.org/10.1021/j150341a001}
  {\bibfield  {journal} {\bibinfo  {journal} {J. Phys. Chem.}\ }\textbf
  {\bibinfo {volume} {36}},\ \bibinfo {pages} {2689} (\bibinfo {year}
  {1932})}\BibitemShut {NoStop}%
\bibitem [{\citenamefont {Zorkot}\ \emph {et~al.}(2016)\citenamefont {Zorkot},
  \citenamefont {Golestanian},\ and\ \citenamefont
  {Bonthuis}}]{zorkot2016power}%
  \BibitemOpen
  \bibfield  {author} {\bibinfo {author} {\bibfnamefont {M.}~\bibnamefont
  {Zorkot}}, \bibinfo {author} {\bibfnamefont {R.}~\bibnamefont
  {Golestanian}},\ and\ \bibinfo {author} {\bibfnamefont {D.~J.}\ \bibnamefont
  {Bonthuis}},\ }\bibfield  {title} {\bibinfo {title} {The power spectrum of
  ionic nanopore currents: the role of ion correlations},\ }\href
  {https://pubs.acs.org/doi/abs/10.1021/acs.nanolett.5b04372} {\bibfield
  {journal} {\bibinfo  {journal} {Nano Lett.}\ }\textbf {\bibinfo {volume}
  {16}},\ \bibinfo {pages} {2205} (\bibinfo {year} {2016})}\BibitemShut
  {NoStop}%
\bibitem [{\citenamefont {D{\'e}mery}\ and\ \citenamefont
  {Dean}(2016)}]{demery2016conductivity}%
  \BibitemOpen
  \bibfield  {author} {\bibinfo {author} {\bibfnamefont {V.}~\bibnamefont
  {D{\'e}mery}}\ and\ \bibinfo {author} {\bibfnamefont {D.~S.}\ \bibnamefont
  {Dean}},\ }\bibfield  {title} {\bibinfo {title} {The conductivity of strong
  electrolytes from stochastic density functional theory},\ }\href
  {https://iopscience.iop.org/article/10.1088/1742-5468/2016/02/023106/meta}
  {\bibfield  {journal} {\bibinfo  {journal} {J. Stat. Mech.: Theory Exp.}\
  }\textbf {\bibinfo {volume} {2016}}\bibinfo  {number} { (2)},\ \bibinfo
  {pages} {023106}}\BibitemShut {NoStop}%
\bibitem [{\citenamefont {Kardar}(2007)}]{kardar-statfield}%
  \BibitemOpen
\bibfield  {number} {  }\bibfield  {author} {\bibinfo {author} {\bibfnamefont
  {M.}~\bibnamefont {Kardar}},\ }\href@noop {} {\emph {\bibinfo {title}
  {Statistical physics of fields}}}\ (\bibinfo  {publisher} {Cambridge
  University Press},\ \bibinfo {year} {2007})\BibitemShut {NoStop}%
\bibitem [{\citenamefont {Dean}(1996)}]{dean96langevin}%
  \BibitemOpen
  \bibfield  {author} {\bibinfo {author} {\bibfnamefont {D.~S.}\ \bibnamefont
  {Dean}},\ }\bibfield  {title} {\bibinfo {title} {{Langevin equation for the
  density of a system of interacting Langevin processes}},\ }\href
  {https://iopscience.iop.org/article/10.1088/0305-4470/29/24/001/meta}
  {\bibfield  {journal} {\bibinfo  {journal} {J. Phys. A}\ }\textbf {\bibinfo
  {volume} {29}},\ \bibinfo {pages} {L613} (\bibinfo {year}
  {1996})}\BibitemShut {NoStop}%
\bibitem [{\citenamefont {Kawasaki}(1994)}]{kawasaki1994}%
  \BibitemOpen
  \bibfield  {author} {\bibinfo {author} {\bibfnamefont {K.}~\bibnamefont
  {Kawasaki}},\ }\bibfield  {title} {\bibinfo {title} {Stochastic model of slow
  dynamics in supercooled liquids and dense colloidal suspensions},\ }\href
  {https://doi.org/10.1016/0378-4371(94)90533-9} {\bibfield  {journal}
  {\bibinfo  {journal} {Physica A}\ }\textbf {\bibinfo {volume} {208}},\
  \bibinfo {pages} {35} (\bibinfo {year} {1994})}\BibitemShut {NoStop}%
\bibitem [{\citenamefont {Frusawa}(2022)}]{frusawa2022electric}%
  \BibitemOpen
  \bibfield  {author} {\bibinfo {author} {\bibfnamefont {H.}~\bibnamefont
  {Frusawa}},\ }\bibfield  {title} {\bibinfo {title} {{Electric-field-induced
  oscillations in ionic fluids: a unified formulation of modified
  Poisson--Nernst--Planck models and its relevance to correlation function
  analysis}},\ }\href
  {https://pubs.rsc.org/en/content/articlehtml/2022/sm/d1sm01811f} {\bibfield
  {journal} {\bibinfo  {journal} {Soft Matter}\ }\textbf {\bibinfo {volume}
  {18}},\ \bibinfo {pages} {4280} (\bibinfo {year} {2022})}\BibitemShut
  {NoStop}%
\bibitem [{\citenamefont {te~Vrugt}\ \emph {et~al.}(2020)\citenamefont
  {te~Vrugt}, \citenamefont {L{\"o}wen},\ and\ \citenamefont
  {Wittkowski}}]{ddftreview}%
  \BibitemOpen
  \bibfield  {author} {\bibinfo {author} {\bibfnamefont {M.}~\bibnamefont
  {te~Vrugt}}, \bibinfo {author} {\bibfnamefont {H.}~\bibnamefont
  {L{\"o}wen}},\ and\ \bibinfo {author} {\bibfnamefont {R.}~\bibnamefont
  {Wittkowski}},\ }\bibfield  {title} {\bibinfo {title} {Classical dynamical
  density functional theory: from fundamentals to applications},\ }\href
  {https://www.tandfonline.com/doi/full/10.1080/00018732.2020.1854965}
  {\bibfield  {journal} {\bibinfo  {journal} {Adv. Phys.}\ }\textbf {\bibinfo
  {volume} {69}},\ \bibinfo {pages} {121} (\bibinfo {year} {2020})}\BibitemShut
  {NoStop}%
\bibitem [{\citenamefont {Grinstein}\ \emph {et~al.}(1990)\citenamefont
  {Grinstein}, \citenamefont {Lee},\ and\ \citenamefont
  {Sachdev}}]{grinstein90conservation}%
  \BibitemOpen
  \bibfield  {author} {\bibinfo {author} {\bibfnamefont {G.}~\bibnamefont
  {Grinstein}}, \bibinfo {author} {\bibfnamefont {D.-H.}\ \bibnamefont {Lee}},\
  and\ \bibinfo {author} {\bibfnamefont {S.}~\bibnamefont {Sachdev}},\
  }\bibfield  {title} {\bibinfo {title} {Conservation laws, anisotropy, and
  ‘‘self-organized criticality’’ in noisy nonequilibrium systems},\
  }\href {https://journals.aps.org/prl/abstract/10.1103/PhysRevLett.64.1927}
  {\bibfield  {journal} {\bibinfo  {journal} {Phys. Rev. Lett.}\ }\textbf
  {\bibinfo {volume} {64}},\ \bibinfo {pages} {1927} (\bibinfo {year}
  {1990})}\BibitemShut {NoStop}%
\bibitem [{\citenamefont {Kardar}\ and\ \citenamefont
  {Golestanian}(1999)}]{kardar99friction}%
  \BibitemOpen
  \bibfield  {author} {\bibinfo {author} {\bibfnamefont {M.}~\bibnamefont
  {Kardar}}\ and\ \bibinfo {author} {\bibfnamefont {R.}~\bibnamefont
  {Golestanian}},\ }\bibfield  {title} {\bibinfo {title} {The ``friction'' of
  vacuum, and other fluctuation-induced forces},\ }\href
  {https://doi.org/10.1103/RevModPhys.71.1233} {\bibfield  {journal} {\bibinfo
  {journal} {Rev. Mod. Phys.}\ }\textbf {\bibinfo {volume} {71}},\ \bibinfo
  {pages} {1233} (\bibinfo {year} {1999})}\BibitemShut {NoStop}%
\bibitem [{\citenamefont {Gambassi}(2009)}]{gambassi2009review}%
  \BibitemOpen
  \bibfield  {author} {\bibinfo {author} {\bibfnamefont {A.}~\bibnamefont
  {Gambassi}},\ }\bibfield  {title} {\bibinfo {title} {{The Casimir effect:
  From quantum to critical fluctuations}},\ }\href
  {https://doi.org/10.1088/1742-6596/161/1/012037} {\bibfield  {journal}
  {\bibinfo  {journal} {J. Phys.: Conf. Ser.}\ }\textbf {\bibinfo {volume}
  {161}},\ \bibinfo {pages} {012037} (\bibinfo {year} {2009})}\BibitemShut
  {NoStop}%
\bibitem [{\citenamefont {Fisher}\ and\ \citenamefont
  {Gennes}(1978)}]{fisher1978wall}%
  \BibitemOpen
  \bibfield  {author} {\bibinfo {author} {\bibfnamefont {M.~E.}\ \bibnamefont
  {Fisher}}\ and\ \bibinfo {author} {\bibfnamefont {P.}~\bibnamefont
  {Gennes}},\ }\bibfield  {title} {\bibinfo {title} {Wall phenomena in a
  critical binary mixture},\ }\href@noop {} {\bibfield  {journal} {\bibinfo
  {journal} {C. R. Acad. Sc. Paris B}\ }\textbf {\bibinfo {volume} {287}},\
  \bibinfo {pages} {207} (\bibinfo {year} {1978})}\BibitemShut {NoStop}%
\bibitem [{\citenamefont {Jackson}(2007)}]{jackson}%
  \BibitemOpen
  \bibfield  {author} {\bibinfo {author} {\bibfnamefont {J.~D.}\ \bibnamefont
  {Jackson}},\ }\href@noop {} {\emph {\bibinfo {title} {Classical
  Electrodynamics}}}\ (\bibinfo  {publisher} {John Wiley \& Sons},\ \bibinfo
  {year} {2007})\BibitemShut {NoStop}%
\bibitem [{\citenamefont {Poncet}\ \emph {et~al.}(2017)\citenamefont {Poncet},
  \citenamefont {B{\'e}nichou}, \citenamefont {D{\'e}mery},\ and\ \citenamefont
  {Oshanin}}]{poncet2017universal}%
  \BibitemOpen
  \bibfield  {author} {\bibinfo {author} {\bibfnamefont {A.}~\bibnamefont
  {Poncet}}, \bibinfo {author} {\bibfnamefont {O.}~\bibnamefont
  {B{\'e}nichou}}, \bibinfo {author} {\bibfnamefont {V.}~\bibnamefont
  {D{\'e}mery}},\ and\ \bibinfo {author} {\bibfnamefont {G.}~\bibnamefont
  {Oshanin}},\ }\bibfield  {title} {\bibinfo {title} {Universal long ranged
  correlations in driven binary mixtures},\ }\href
  {https://journals.aps.org/prl/abstract/10.1103/PhysRevLett.118.118002}
  {\bibfield  {journal} {\bibinfo  {journal} {Phys. Rev. Lett.}\ }\textbf
  {\bibinfo {volume} {118}},\ \bibinfo {pages} {118002} (\bibinfo {year}
  {2017})}\BibitemShut {NoStop}%
\bibitem [{\citenamefont {Richter}\ \emph {et~al.}(2020)\citenamefont
  {Richter}, \citenamefont {\ifmmode~\dot{Z}\else \.{Z}\fi{}uk}, \citenamefont
  {Szymczak}, \citenamefont {Paczesny}, \citenamefont {Bk{a}k}, \citenamefont
  {Szymborski}, \citenamefont {Garstecki}, \citenamefont {Stone}, \citenamefont
  {Ho\l{}yst},\ and\ \citenamefont {Drummond}}]{stoneAC}%
  \BibitemOpen
  \bibfield  {author} {\bibinfo {author} {\bibfnamefont {L.}~\bibnamefont
  {Richter}}, \bibinfo {author} {\bibfnamefont {P.~J.}\ \bibnamefont
  {\ifmmode~\dot{Z}\else \.{Z}\fi{}uk}}, \bibinfo {author} {\bibfnamefont
  {P.}~\bibnamefont {Szymczak}}, \bibinfo {author} {\bibfnamefont
  {J.}~\bibnamefont {Paczesny}}, \bibinfo {author} {\bibfnamefont {K.~M.}\
  \bibnamefont {Bk{a}k}}, \bibinfo {author} {\bibfnamefont {T.}~\bibnamefont
  {Szymborski}}, \bibinfo {author} {\bibfnamefont {P.}~\bibnamefont
  {Garstecki}}, \bibinfo {author} {\bibfnamefont {H.~A.}\ \bibnamefont
  {Stone}}, \bibinfo {author} {\bibfnamefont {R.}~\bibnamefont {Ho\l{}yst}},\
  and\ \bibinfo {author} {\bibfnamefont {C.}~\bibnamefont {Drummond}},\
  }\bibfield  {title} {\bibinfo {title} {Ions in an ac electric field: Strong
  long-range repulsion between oppositely charged surfaces},\ }\href
  {https://doi.org/10.1103/PhysRevLett.125.056001} {\bibfield  {journal}
  {\bibinfo  {journal} {Phys. Rev. Lett.}\ }\textbf {\bibinfo {volume} {125}},\
  \bibinfo {pages} {056001} (\bibinfo {year} {2020})}\BibitemShut {NoStop}%
\bibitem [{\citenamefont {Israelachvili}(2011)}]{israelachvili}%
  \BibitemOpen
  \bibfield  {author} {\bibinfo {author} {\bibfnamefont {J.~N.}\ \bibnamefont
  {Israelachvili}},\ }\href@noop {} {\emph {\bibinfo {title} {Intermolecular
  and surface forces}}}\ (\bibinfo  {publisher} {Academic press},\ \bibinfo
  {year} {2011})\BibitemShut {NoStop}%
\bibitem [{\citenamefont {Amrei}\ \emph {et~al.}(2018)\citenamefont {Amrei},
  \citenamefont {Bukosky}, \citenamefont {Rader}, \citenamefont {Ristenpart},\
  and\ \citenamefont {Miller}}]{amrei2018oscillating}%
  \BibitemOpen
  \bibfield  {author} {\bibinfo {author} {\bibfnamefont {S.~H.}\ \bibnamefont
  {Amrei}}, \bibinfo {author} {\bibfnamefont {S.~C.}\ \bibnamefont {Bukosky}},
  \bibinfo {author} {\bibfnamefont {S.~P.}\ \bibnamefont {Rader}}, \bibinfo
  {author} {\bibfnamefont {W.~D.}\ \bibnamefont {Ristenpart}},\ and\ \bibinfo
  {author} {\bibfnamefont {G.~H.}\ \bibnamefont {Miller}},\ }\bibfield  {title}
  {\bibinfo {title} {Oscillating electric fields in liquids create a long-range
  steady field},\ }\href
  {https://journals.aps.org/prl/abstract/10.1103/PhysRevLett.121.185504}
  {\bibfield  {journal} {\bibinfo  {journal} {Phys. Rev. Lett.}\ }\textbf
  {\bibinfo {volume} {121}},\ \bibinfo {pages} {185504} (\bibinfo {year}
  {2018})}\BibitemShut {NoStop}%
\bibitem [{\citenamefont {Levine}\ and\ \citenamefont
  {Rappel}(2013)}]{levine2013}%
  \BibitemOpen
  \bibfield  {author} {\bibinfo {author} {\bibfnamefont {H.}~\bibnamefont
  {Levine}}\ and\ \bibinfo {author} {\bibfnamefont {W.-J.}\ \bibnamefont
  {Rappel}},\ }\bibfield  {title} {\bibinfo {title} {The physics of eukaryotic
  chemotaxis},\ }\href {https://doi.org/10.1063/PT.3.1884} {\bibfield
  {journal} {\bibinfo  {journal} {Phys. Today}\ }\textbf {\bibinfo {volume}
  {66}},\ \bibinfo {pages} {24} (\bibinfo {year} {2013})}\BibitemShut {NoStop}%
\bibitem [{\citenamefont {Camley}(2018)}]{camley2018collective}%
  \BibitemOpen
  \bibfield  {author} {\bibinfo {author} {\bibfnamefont {B.~A.}\ \bibnamefont
  {Camley}},\ }\bibfield  {title} {\bibinfo {title} {Collective gradient
  sensing and chemotaxis: modeling and recent developments},\ }\href
  {https://iopscience.iop.org/article/10.1088/1361-648X/aabd9f/meta} {\bibfield
   {journal} {\bibinfo  {journal} {J. Phys. Condens. Matter}\ }\textbf
  {\bibinfo {volume} {30}},\ \bibinfo {pages} {223001} (\bibinfo {year}
  {2018})}\BibitemShut {NoStop}%
\bibitem [{\citenamefont {Bagorda}\ and\ \citenamefont
  {Parent}(2008)}]{chemotaxis-glance}%
  \BibitemOpen
  \bibfield  {author} {\bibinfo {author} {\bibfnamefont {A.}~\bibnamefont
  {Bagorda}}\ and\ \bibinfo {author} {\bibfnamefont {C.~A.}\ \bibnamefont
  {Parent}},\ }\bibfield  {title} {\bibinfo {title} {Eukaryotic chemotaxis at a
  glance},\ }\href
  {https://journals.biologists.com/jcs/article/121/16/2621/30151/Eukaryotic-chemotaxis-at-a-glance}
  {\bibfield  {journal} {\bibinfo  {journal} {J. Cell Sci.}\ }\textbf {\bibinfo
  {volume} {121}},\ \bibinfo {pages} {2621} (\bibinfo {year}
  {2008})}\BibitemShut {NoStop}%
\bibitem [{\citenamefont {Berg}(2000)}]{berg2000motile}%
  \BibitemOpen
  \bibfield  {author} {\bibinfo {author} {\bibfnamefont {H.}~\bibnamefont
  {Berg}},\ }\bibfield  {title} {\bibinfo {title} {Motile behavior of
  bacteria},\ }\href@noop {} {\bibfield  {journal} {\bibinfo  {journal} {Phys.
  today}\ } (\bibinfo {year} {2000})}\BibitemShut {NoStop}%
\bibitem [{\citenamefont {Van~Haastert}\ and\ \citenamefont
  {Devreotes}(2004)}]{van2004chemotaxis}%
  \BibitemOpen
  \bibfield  {author} {\bibinfo {author} {\bibfnamefont {P.~J.}\ \bibnamefont
  {Van~Haastert}}\ and\ \bibinfo {author} {\bibfnamefont {P.~N.}\ \bibnamefont
  {Devreotes}},\ }\bibfield  {title} {\bibinfo {title} {Chemotaxis: signalling
  the way forward},\ }\href
  {https://www.nature.com/articles/nrm1435?message=remove&ref=theblueish.com/web}
  {\bibfield  {journal} {\bibinfo  {journal} {Nat. Rev. Mol. Cell Biol.}\
  }\textbf {\bibinfo {volume} {5}},\ \bibinfo {pages} {626} (\bibinfo {year}
  {2004})}\BibitemShut {NoStop}%
\bibitem [{\citenamefont {Dey}\ \emph {et~al.}(2014)\citenamefont {Dey},
  \citenamefont {Das}, \citenamefont {Poyton}, \citenamefont {Sengupta},
  \citenamefont {Butler}, \citenamefont {Cremer},\ and\ \citenamefont
  {Sen}}]{dey14}%
  \BibitemOpen
  \bibfield  {author} {\bibinfo {author} {\bibfnamefont {K.~K.}\ \bibnamefont
  {Dey}}, \bibinfo {author} {\bibfnamefont {S.}~\bibnamefont {Das}}, \bibinfo
  {author} {\bibfnamefont {M.~F.}\ \bibnamefont {Poyton}}, \bibinfo {author}
  {\bibfnamefont {S.}~\bibnamefont {Sengupta}}, \bibinfo {author}
  {\bibfnamefont {P.~J.}\ \bibnamefont {Butler}}, \bibinfo {author}
  {\bibfnamefont {P.~S.}\ \bibnamefont {Cremer}},\ and\ \bibinfo {author}
  {\bibfnamefont {A.}~\bibnamefont {Sen}},\ }\bibfield  {title} {\bibinfo
  {title} {Chemotactic separation of enzymes},\ }\href
  {https://pubs.acs.org/doi/abs/10.1021/nn504418u} {\bibfield  {journal}
  {\bibinfo  {journal} {ACS Nano}\ }\textbf {\bibinfo {volume} {8}},\ \bibinfo
  {pages} {11941} (\bibinfo {year} {2014})}\BibitemShut {NoStop}%
\bibitem [{\citenamefont {{Agudo-Canalejo}}\ \emph {et~al.}(2018)\citenamefont
  {{Agudo-Canalejo}}, \citenamefont {Illien},\ and\ \citenamefont
  {Golestanian}}]{agudo2018}%
  \BibitemOpen
  \bibfield  {author} {\bibinfo {author} {\bibfnamefont {J.}~\bibnamefont
  {{Agudo-Canalejo}}}, \bibinfo {author} {\bibfnamefont {P.}~\bibnamefont
  {Illien}},\ and\ \bibinfo {author} {\bibfnamefont {R.}~\bibnamefont
  {Golestanian}},\ }\bibfield  {title} {\bibinfo {title} {Phoresis and
  {{Enhanced Diffusion Compete}} in {{Enzyme Chemotaxis}}},\ }\href
  {https://doi.org/10.1021/acs.nanolett.8b00717} {\bibfield  {journal}
  {\bibinfo  {journal} {Nano Lett.}\ }\textbf {\bibinfo {volume} {18}},\
  \bibinfo {pages} {2711} (\bibinfo {year} {2018})}\BibitemShut {NoStop}%
\bibitem [{\citenamefont {Thakur}\ and\ \citenamefont
  {Kapral}(2012)}]{Kapral2012}%
  \BibitemOpen
  \bibfield  {author} {\bibinfo {author} {\bibfnamefont {S.}~\bibnamefont
  {Thakur}}\ and\ \bibinfo {author} {\bibfnamefont {R.}~\bibnamefont
  {Kapral}},\ }\bibfield  {title} {\bibinfo {title} {Collective dynamics of
  self-propelled sphere-dimer motors},\ }\href
  {https://journals.aps.org/pre/abstract/10.1103/PhysRevE.85.026121} {\bibfield
   {journal} {\bibinfo  {journal} {Phys. Rev. E}\ }\textbf {\bibinfo {volume}
  {85}},\ \bibinfo {pages} {026121} (\bibinfo {year} {2012})}\BibitemShut
  {NoStop}%
\bibitem [{\citenamefont {Soto}\ and\ \citenamefont
  {Golestanian}(2014)}]{Soto2014}%
  \BibitemOpen
  \bibfield  {author} {\bibinfo {author} {\bibfnamefont {R.}~\bibnamefont
  {Soto}}\ and\ \bibinfo {author} {\bibfnamefont {R.}~\bibnamefont
  {Golestanian}},\ }\bibfield  {title} {\bibinfo {title} {Run-and-tumble
  dynamics in a crowded environment: Persistent exclusion process for
  swimmers},\ }\href
  {https://journals.aps.org/pre/abstract/10.1103/PhysRevE.89.012706} {\bibfield
   {journal} {\bibinfo  {journal} {Phys. Rev. E}\ }\textbf {\bibinfo {volume}
  {89}},\ \bibinfo {pages} {012706} (\bibinfo {year} {2014})}\BibitemShut
  {NoStop}%
\bibitem [{\citenamefont {Illien}\ \emph {et~al.}(2017)\citenamefont {Illien},
  \citenamefont {Golestanian},\ and\ \citenamefont {Sen}}]{illien2017}%
  \BibitemOpen
  \bibfield  {author} {\bibinfo {author} {\bibfnamefont {P.}~\bibnamefont
  {Illien}}, \bibinfo {author} {\bibfnamefont {R.}~\bibnamefont
  {Golestanian}},\ and\ \bibinfo {author} {\bibfnamefont {A.}~\bibnamefont
  {Sen}},\ }\bibfield  {title} {\bibinfo {title} {`{{Fuelled}}' motion:
  Phoretic motility and collective behaviour of active colloids},\ }\href
  {https://doi.org/10.1039/C7CS00087A} {\bibfield  {journal} {\bibinfo
  {journal} {Chem. Soc. Rev.}\ }\textbf {\bibinfo {volume} {46}},\ \bibinfo
  {pages} {5508} (\bibinfo {year} {2017})}\BibitemShut {NoStop}%
\bibitem [{\citenamefont {Stark}(2018)}]{stark2018}%
  \BibitemOpen
  \bibfield  {author} {\bibinfo {author} {\bibfnamefont {H.}~\bibnamefont
  {Stark}},\ }\bibfield  {title} {\bibinfo {title} {Artificial {{Chemotaxis}}
  of {{Self}}-{{Phoretic Active Colloids}}: {{Collective Behavior}}},\ }\href
  {https://doi.org/10.1021/acs.accounts.8b00259} {\bibfield  {journal}
  {\bibinfo  {journal} {Acc. Chem. Res.}\ }\textbf {\bibinfo {volume} {51}},\
  \bibinfo {pages} {2681} (\bibinfo {year} {2018})}\BibitemShut {NoStop}%
\bibitem [{\citenamefont {Friedl}\ and\ \citenamefont
  {Gilmour}(2009)}]{friedl2009}%
  \BibitemOpen
  \bibfield  {author} {\bibinfo {author} {\bibfnamefont {P.}~\bibnamefont
  {Friedl}}\ and\ \bibinfo {author} {\bibfnamefont {D.}~\bibnamefont
  {Gilmour}},\ }\bibfield  {title} {\bibinfo {title} {Collective cell migration
  in morphogenesis, regeneration and cancer},\ }\href
  {https://doi.org/10.1038/nrm2720} {\bibfield  {journal} {\bibinfo  {journal}
  {Nat. Rev. Mol. Cell Biol.}\ }\textbf {\bibinfo {volume} {10}},\ \bibinfo
  {pages} {445} (\bibinfo {year} {2009})}\BibitemShut {NoStop}%
\bibitem [{\citenamefont {Gini{\=u}nait{\.e}}\ \emph
  {et~al.}(2020)\citenamefont {Gini{\=u}nait{\.e}}, \citenamefont {Baker},
  \citenamefont {Kulesa},\ and\ \citenamefont
  {Maini}}]{giniunaite2020modelling}%
  \BibitemOpen
  \bibfield  {author} {\bibinfo {author} {\bibfnamefont {R.}~\bibnamefont
  {Gini{\=u}nait{\.e}}}, \bibinfo {author} {\bibfnamefont {R.~E.}\ \bibnamefont
  {Baker}}, \bibinfo {author} {\bibfnamefont {P.~M.}\ \bibnamefont {Kulesa}},\
  and\ \bibinfo {author} {\bibfnamefont {P.~K.}\ \bibnamefont {Maini}},\
  }\bibfield  {title} {\bibinfo {title} {Modelling collective cell migration:
  neural crest as a model paradigm},\ }\href
  {https://link.springer.com/article/10.1007/s00285-019-01436-2} {\bibfield
  {journal} {\bibinfo  {journal} {J. Math. Biol.}\ }\textbf {\bibinfo {volume}
  {80}},\ \bibinfo {pages} {481} (\bibinfo {year} {2020})}\BibitemShut
  {NoStop}%
\bibitem [{\citenamefont {Hogan}(1999)}]{hogan1999}%
  \BibitemOpen
  \bibfield  {author} {\bibinfo {author} {\bibfnamefont {B.~L.}\ \bibnamefont
  {Hogan}},\ }\bibfield  {title} {\bibinfo {title} {Morphogenesis},\ }\href
  {https://doi.org/10.1016/S0092-8674(00)80562-0} {\bibfield  {journal}
  {\bibinfo  {journal} {Cell}\ }\textbf {\bibinfo {volume} {96}},\ \bibinfo
  {pages} {225} (\bibinfo {year} {1999})}\BibitemShut {NoStop}%
\bibitem [{\citenamefont {Crick}(1970)}]{crick1970}%
  \BibitemOpen
  \bibfield  {author} {\bibinfo {author} {\bibfnamefont {F.}~\bibnamefont
  {Crick}},\ }\bibfield  {title} {\bibinfo {title} {Diffusion in
  {{Embryogenesis}}},\ }\href {https://doi.org/10.1038/225420a0} {\bibfield
  {journal} {\bibinfo  {journal} {Nature}\ }\textbf {\bibinfo {volume} {225}},\
  \bibinfo {pages} {420} (\bibinfo {year} {1970})}\BibitemShut {NoStop}%
\bibitem [{\citenamefont {De~Oliveira}\ \emph {et~al.}(2016)\citenamefont
  {De~Oliveira}, \citenamefont {Rosowski},\ and\ \citenamefont
  {Huttenlocher}}]{wound-review}%
  \BibitemOpen
  \bibfield  {author} {\bibinfo {author} {\bibfnamefont {S.}~\bibnamefont
  {De~Oliveira}}, \bibinfo {author} {\bibfnamefont {E.~E.}\ \bibnamefont
  {Rosowski}},\ and\ \bibinfo {author} {\bibfnamefont {A.}~\bibnamefont
  {Huttenlocher}},\ }\bibfield  {title} {\bibinfo {title} {Neutrophil migration
  in infection and wound repair: going forward in reverse},\ }\href
  {https://www.nature.com/articles/nri.2016.49} {\bibfield  {journal} {\bibinfo
   {journal} {Nat. Rev. Immunol.}\ }\textbf {\bibinfo {volume} {16}},\ \bibinfo
  {pages} {378} (\bibinfo {year} {2016})}\BibitemShut {NoStop}%
\bibitem [{\citenamefont {Petri}\ and\ \citenamefont
  {Sanz}(2018)}]{petri2018neutrophil}%
  \BibitemOpen
  \bibfield  {author} {\bibinfo {author} {\bibfnamefont {B.}~\bibnamefont
  {Petri}}\ and\ \bibinfo {author} {\bibfnamefont {M.-J.}\ \bibnamefont
  {Sanz}},\ }\bibfield  {title} {\bibinfo {title} {Neutrophil chemotaxis},\
  }\href {https://link.springer.com/article/10.1007/s00441-017-2776-8}
  {\bibfield  {journal} {\bibinfo  {journal} {Cell Tissue Res.}\ }\textbf
  {\bibinfo {volume} {371}},\ \bibinfo {pages} {425} (\bibinfo {year}
  {2018})}\BibitemShut {NoStop}%
\bibitem [{\citenamefont {Schneider}\ \emph {et~al.}(2010)\citenamefont
  {Schneider}, \citenamefont {Cammer}, \citenamefont {Lehman}, \citenamefont
  {Nielsen}, \citenamefont {Guerra}, \citenamefont {Veland}, \citenamefont
  {Stock}, \citenamefont {Hoffmann}, \citenamefont {Yoder}, \citenamefont
  {Schwab} \emph {et~al.}}]{fibroblast-wound}%
  \BibitemOpen
  \bibfield  {author} {\bibinfo {author} {\bibfnamefont {L.}~\bibnamefont
  {Schneider}}, \bibinfo {author} {\bibfnamefont {M.}~\bibnamefont {Cammer}},
  \bibinfo {author} {\bibfnamefont {J.}~\bibnamefont {Lehman}}, \bibinfo
  {author} {\bibfnamefont {S.~K.}\ \bibnamefont {Nielsen}}, \bibinfo {author}
  {\bibfnamefont {C.~F.}\ \bibnamefont {Guerra}}, \bibinfo {author}
  {\bibfnamefont {I.~R.}\ \bibnamefont {Veland}}, \bibinfo {author}
  {\bibfnamefont {C.}~\bibnamefont {Stock}}, \bibinfo {author} {\bibfnamefont
  {E.~K.}\ \bibnamefont {Hoffmann}}, \bibinfo {author} {\bibfnamefont {B.~K.}\
  \bibnamefont {Yoder}}, \bibinfo {author} {\bibfnamefont {A.}~\bibnamefont
  {Schwab}}, \emph {et~al.},\ }\bibfield  {title} {\bibinfo {title}
  {Directional cell migration and chemotaxis in wound healing response to
  pdgf-aa are coordinated by the primary cilium in fibroblasts},\ }\href@noop
  {} {\bibfield  {journal} {\bibinfo  {journal} {Cell. Physiol. Biochem.}\
  }\textbf {\bibinfo {volume} {25}},\ \bibinfo {pages} {279} (\bibinfo {year}
  {2010})}\BibitemShut {NoStop}%
\bibitem [{\citenamefont {Roussos}\ \emph {et~al.}(2011)\citenamefont
  {Roussos}, \citenamefont {Condeelis},\ and\ \citenamefont
  {Patsialou}}]{roussos2011}%
  \BibitemOpen
  \bibfield  {author} {\bibinfo {author} {\bibfnamefont {E.~T.}\ \bibnamefont
  {Roussos}}, \bibinfo {author} {\bibfnamefont {J.~S.}\ \bibnamefont
  {Condeelis}},\ and\ \bibinfo {author} {\bibfnamefont {A.}~\bibnamefont
  {Patsialou}},\ }\bibfield  {title} {\bibinfo {title} {Chemotaxis in cancer},\
  }\href {https://doi.org/10.1038/nrc3078} {\bibfield  {journal} {\bibinfo
  {journal} {Nat. Rev. Cancer}\ }\textbf {\bibinfo {volume} {11}},\ \bibinfo
  {pages} {573} (\bibinfo {year} {2011})}\BibitemShut {NoStop}%
\bibitem [{\citenamefont {Hanahan}\ and\ \citenamefont
  {Weinberg}(2011)}]{hanahan2011}%
  \BibitemOpen
  \bibfield  {author} {\bibinfo {author} {\bibfnamefont {D.}~\bibnamefont
  {Hanahan}}\ and\ \bibinfo {author} {\bibfnamefont {R.~A.}\ \bibnamefont
  {Weinberg}},\ }\bibfield  {title} {\bibinfo {title} {Hallmarks of {{Cancer}}:
  {{The Next Generation}}},\ }\href
  {https://doi.org/10.1016/j.cell.2011.02.013} {\bibfield  {journal} {\bibinfo
  {journal} {Cell}\ }\textbf {\bibinfo {volume} {144}},\ \bibinfo {pages} {646}
  (\bibinfo {year} {2011})}\BibitemShut {NoStop}%
\bibitem [{\citenamefont {Keller}\ and\ \citenamefont
  {Segel}(1971)}]{keller1971}%
  \BibitemOpen
  \bibfield  {author} {\bibinfo {author} {\bibfnamefont {E.~F.}\ \bibnamefont
  {Keller}}\ and\ \bibinfo {author} {\bibfnamefont {L.~A.}\ \bibnamefont
  {Segel}},\ }\bibfield  {title} {\bibinfo {title} {Model for chemotaxis},\
  }\href {https://doi.org/10.1016/0022-5193(71)90050-6} {\bibfield  {journal}
  {\bibinfo  {journal} {J. Theor. Biol.}\ }\textbf {\bibinfo {volume} {30}},\
  \bibinfo {pages} {225} (\bibinfo {year} {1971})}\BibitemShut {NoStop}%
\bibitem [{\citenamefont {Patlak}(1953)}]{patlak1953random}%
  \BibitemOpen
  \bibfield  {author} {\bibinfo {author} {\bibfnamefont {C.~S.}\ \bibnamefont
  {Patlak}},\ }\bibfield  {title} {\bibinfo {title} {Random walk with
  persistence and external bias},\ }\href
  {https://link.springer.com/article/10.1007/BF02476407} {\bibfield  {journal}
  {\bibinfo  {journal} {B. Math. Biophys.}\ }\textbf {\bibinfo {volume} {15}},\
  \bibinfo {pages} {311} (\bibinfo {year} {1953})}\BibitemShut {NoStop}%
\bibitem [{\citenamefont {Chavanis}(2010)}]{chavanis2010stochasticKS}%
  \BibitemOpen
  \bibfield  {author} {\bibinfo {author} {\bibfnamefont {P.-H.}\ \bibnamefont
  {Chavanis}},\ }\bibfield  {title} {\bibinfo {title} {A stochastic
  {{Keller}}\textendash{{Segel}} model of chemotaxis},\ }\href
  {https://doi.org/10.1016/j.cnsns.2008.09.002} {\bibfield  {journal} {\bibinfo
   {journal} {Commun. Nonlinear Sci. Numer. Simul.}\ }\textbf {\bibinfo
  {volume} {15}},\ \bibinfo {pages} {60} (\bibinfo {year} {2010})}\BibitemShut
  {NoStop}%
\bibitem [{\citenamefont {Newman}\ and\ \citenamefont
  {Grima}(2004)}]{grima-manybody}%
  \BibitemOpen
  \bibfield  {author} {\bibinfo {author} {\bibfnamefont {T.~J.}\ \bibnamefont
  {Newman}}\ and\ \bibinfo {author} {\bibfnamefont {R.}~\bibnamefont {Grima}},\
  }\bibfield  {title} {\bibinfo {title} {Many-body theory of chemotactic
  cell-cell interactions},\ }\href {https://doi.org/10.1103/PhysRevE.70.051916}
  {\bibfield  {journal} {\bibinfo  {journal} {Phys. Rev. E}\ }\textbf {\bibinfo
  {volume} {70}},\ \bibinfo {pages} {051916} (\bibinfo {year}
  {2004})}\BibitemShut {NoStop}%
\bibitem [{\citenamefont {Hillen}\ and\ \citenamefont
  {Painter}(2009)}]{hillen2009}%
  \BibitemOpen
  \bibfield  {author} {\bibinfo {author} {\bibfnamefont {T.}~\bibnamefont
  {Hillen}}\ and\ \bibinfo {author} {\bibfnamefont {K.~J.}\ \bibnamefont
  {Painter}},\ }\bibfield  {title} {\bibinfo {title} {A user's guide to {{PDE}}
  models for chemotaxis},\ }\href {https://doi.org/10.1007/s00285-008-0201-3}
  {\bibfield  {journal} {\bibinfo  {journal} {J. Math. Biol.}\ }\textbf
  {\bibinfo {volume} {58}},\ \bibinfo {pages} {183} (\bibinfo {year}
  {2009})}\BibitemShut {NoStop}%
\bibitem [{\citenamefont {H{\"o}fer}\ \emph {et~al.}(1995)\citenamefont
  {H{\"o}fer}, \citenamefont {Sherratt},\ and\ \citenamefont
  {Maini}}]{diffusion-difference1}%
  \BibitemOpen
  \bibfield  {author} {\bibinfo {author} {\bibfnamefont {T.}~\bibnamefont
  {H{\"o}fer}}, \bibinfo {author} {\bibfnamefont {J.~A.}\ \bibnamefont
  {Sherratt}},\ and\ \bibinfo {author} {\bibfnamefont {P.~K.}\ \bibnamefont
  {Maini}},\ }\bibfield  {title} {\bibinfo {title} {Cellular pattern formation
  during dictyostelium aggregation},\ }\href
  {https://www.sciencedirect.com/science/article/abs/pii/016727899500075F}
  {\bibfield  {journal} {\bibinfo  {journal} {Physica D}\ }\textbf {\bibinfo
  {volume} {85}},\ \bibinfo {pages} {425} (\bibinfo {year} {1995})}\BibitemShut
  {NoStop}%
\bibitem [{\citenamefont {Luca}\ \emph {et~al.}(2003)\citenamefont {Luca},
  \citenamefont {Chavez-Ross}, \citenamefont {Edelstein-Keshet},\ and\
  \citenamefont {Mogilner}}]{diffusion-difference2}%
  \BibitemOpen
  \bibfield  {author} {\bibinfo {author} {\bibfnamefont {M.}~\bibnamefont
  {Luca}}, \bibinfo {author} {\bibfnamefont {A.}~\bibnamefont {Chavez-Ross}},
  \bibinfo {author} {\bibfnamefont {L.}~\bibnamefont {Edelstein-Keshet}},\ and\
  \bibinfo {author} {\bibfnamefont {A.}~\bibnamefont {Mogilner}},\ }\bibfield
  {title} {\bibinfo {title} {Chemotactic signaling, microglia, and
  alzheimer’s disease senile plaques: Is there a connection?},\ }\href
  {https://link.springer.com/article/10.1016/S0092-8240(03)00030-2} {\bibfield
  {journal} {\bibinfo  {journal} {Bull. Math. Biol.}\ }\textbf {\bibinfo
  {volume} {65}},\ \bibinfo {pages} {693} (\bibinfo {year} {2003})}\BibitemShut
  {NoStop}%
\bibitem [{\citenamefont {Chavanis}\ and\ \citenamefont
  {Sire}(2007)}]{chavanis-kineticchemo}%
  \BibitemOpen
  \bibfield  {author} {\bibinfo {author} {\bibfnamefont {P.-H.}\ \bibnamefont
  {Chavanis}}\ and\ \bibinfo {author} {\bibfnamefont {C.}~\bibnamefont
  {Sire}},\ }\bibfield  {title} {\bibinfo {title} {Kinetic and hydrodynamic
  models of chemotactic aggregation},\ }\href
  {https://doi.org/10.1016/j.physa.2007.05.069} {\bibfield  {journal} {\bibinfo
   {journal} {Physica A}\ }\textbf {\bibinfo {volume} {384}},\ \bibinfo {pages}
  {199} (\bibinfo {year} {2007})}\BibitemShut {NoStop}%
\bibitem [{\citenamefont {J{\"a}ger}\ and\ \citenamefont
  {Luckhaus}(1992)}]{jager1992explosions}%
  \BibitemOpen
  \bibfield  {author} {\bibinfo {author} {\bibfnamefont {W.}~\bibnamefont
  {J{\"a}ger}}\ and\ \bibinfo {author} {\bibfnamefont {S.}~\bibnamefont
  {Luckhaus}},\ }\bibfield  {title} {\bibinfo {title} {On explosions of
  solutions to a system of partial differential equations modelling
  chemotaxis},\ }\href
  {https://www.ams.org/journals/tran/1992-329-02/S0002-9947-1992-1046835-6/}
  {\bibfield  {journal} {\bibinfo  {journal} {Trans. Am. Math. Soc.}\ }\textbf
  {\bibinfo {volume} {329}},\ \bibinfo {pages} {819} (\bibinfo {year}
  {1992})}\BibitemShut {NoStop}%
\bibitem [{\citenamefont {Yeo}\ \emph {et~al.}(2018)\citenamefont {Yeo},
  \citenamefont {Lee}, \citenamefont {Shin}, \citenamefont {An}, \citenamefont
  {Cho},\ and\ \citenamefont {Kim}}]{yeo2018}%
  \BibitemOpen
  \bibfield  {author} {\bibinfo {author} {\bibfnamefont {S.-Y.}\ \bibnamefont
  {Yeo}}, \bibinfo {author} {\bibfnamefont {K.-W.}\ \bibnamefont {Lee}},
  \bibinfo {author} {\bibfnamefont {D.}~\bibnamefont {Shin}}, \bibinfo {author}
  {\bibfnamefont {S.}~\bibnamefont {An}}, \bibinfo {author} {\bibfnamefont
  {K.-H.}\ \bibnamefont {Cho}},\ and\ \bibinfo {author} {\bibfnamefont {S.-H.}\
  \bibnamefont {Kim}},\ }\bibfield  {title} {\bibinfo {title} {A positive
  feedback loop bi-stably activates fibroblasts},\ }\href
  {https://www.nature.com/articles/s41467-018-05274-6} {\bibfield  {journal}
  {\bibinfo  {journal} {Nat. Commun.}\ }\textbf {\bibinfo {volume} {9}},\
  \bibinfo {pages} {1} (\bibinfo {year} {2018})}\BibitemShut {NoStop}%
\bibitem [{\citenamefont {Kardar}\ \emph {et~al.}(1986)\citenamefont {Kardar},
  \citenamefont {Parisi},\ and\ \citenamefont {Zhang}}]{kardar1986}%
  \BibitemOpen
  \bibfield  {author} {\bibinfo {author} {\bibfnamefont {M.}~\bibnamefont
  {Kardar}}, \bibinfo {author} {\bibfnamefont {G.}~\bibnamefont {Parisi}},\
  and\ \bibinfo {author} {\bibfnamefont {Y.-C.}\ \bibnamefont {Zhang}},\
  }\bibfield  {title} {\bibinfo {title} {Dynamic scaling of growing
  interfaces},\ }\href
  {https://journals.aps.org/prl/abstract/10.1103/PhysRevLett.56.889} {\bibfield
   {journal} {\bibinfo  {journal} {Phys. Rev. Lett.}\ }\textbf {\bibinfo
  {volume} {56}},\ \bibinfo {pages} {889} (\bibinfo {year} {1986})}\BibitemShut
  {NoStop}%
\bibitem [{\citenamefont {Medina}\ \emph {et~al.}(1989)\citenamefont {Medina},
  \citenamefont {Hwa}, \citenamefont {Kardar},\ and\ \citenamefont
  {Zhang}}]{medina1989}%
  \BibitemOpen
  \bibfield  {author} {\bibinfo {author} {\bibfnamefont {E.}~\bibnamefont
  {Medina}}, \bibinfo {author} {\bibfnamefont {T.}~\bibnamefont {Hwa}},
  \bibinfo {author} {\bibfnamefont {M.}~\bibnamefont {Kardar}},\ and\ \bibinfo
  {author} {\bibfnamefont {Y.-C.}\ \bibnamefont {Zhang}},\ }\bibfield  {title}
  {\bibinfo {title} {Burgers equation with correlated noise:
  {{Renormalization}}-group analysis and applications to directed polymers and
  interface growth},\ }\href {https://doi.org/10.1103/PhysRevA.39.3053}
  {\bibfield  {journal} {\bibinfo  {journal} {Phys. Rev. A}\ }\textbf {\bibinfo
  {volume} {39}},\ \bibinfo {pages} {3053} (\bibinfo {year}
  {1989})}\BibitemShut {NoStop}%
\bibitem [{\citenamefont {Barab{\'a}si}\ and\ \citenamefont
  {Stanley}(1995)}]{barabasi}%
  \BibitemOpen
  \bibfield  {author} {\bibinfo {author} {\bibfnamefont {A.-L.}\ \bibnamefont
  {Barab{\'a}si}}\ and\ \bibinfo {author} {\bibfnamefont {H.~E.}\ \bibnamefont
  {Stanley}},\ }\href@noop {} {\emph {\bibinfo {title} {Fractal concepts in
  surface growth}}}\ (\bibinfo  {publisher} {Cambridge University Press},\
  \bibinfo {year} {1995})\BibitemShut {NoStop}%
\bibitem [{\citenamefont {Servant}\ \emph {et~al.}(1999)\citenamefont
  {Servant}, \citenamefont {Weiner}, \citenamefont {Neptune}, \citenamefont
  {Sedat},\ and\ \citenamefont {Bourne}}]{servant-dynamics99}%
  \BibitemOpen
  \bibfield  {author} {\bibinfo {author} {\bibfnamefont {G.}~\bibnamefont
  {Servant}}, \bibinfo {author} {\bibfnamefont {O.~D.}\ \bibnamefont {Weiner}},
  \bibinfo {author} {\bibfnamefont {E.~R.}\ \bibnamefont {Neptune}}, \bibinfo
  {author} {\bibfnamefont {J.~W.}\ \bibnamefont {Sedat}},\ and\ \bibinfo
  {author} {\bibfnamefont {H.~R.}\ \bibnamefont {Bourne}},\ }\bibfield  {title}
  {\bibinfo {title} {Dynamics of a chemoattractant receptor in living
  neutrophils during chemotaxis},\ }\href
  {https://www.molbiolcell.org/doi/full/10.1091/mbc.10.4.1163} {\bibfield
  {journal} {\bibinfo  {journal} {Mol. Biol. Cell}\ }\textbf {\bibinfo {volume}
  {10}},\ \bibinfo {pages} {1163} (\bibinfo {year} {1999})}\BibitemShut
  {NoStop}%
\bibitem [{\citenamefont {Devreotes}\ and\ \citenamefont
  {Janetopoulos}(2003)}]{devreotes2003eukaryotic}%
  \BibitemOpen
  \bibfield  {author} {\bibinfo {author} {\bibfnamefont {P.}~\bibnamefont
  {Devreotes}}\ and\ \bibinfo {author} {\bibfnamefont {C.}~\bibnamefont
  {Janetopoulos}},\ }\bibfield  {title} {\bibinfo {title} {Eukaryotic
  chemotaxis: distinctions between directional sensing and polarization},\
  }\href {https://www.jbc.org/article/S0021-9258(20)73334-X/fulltext}
  {\bibfield  {journal} {\bibinfo  {journal} {J. Biol. Chem.}\ }\textbf
  {\bibinfo {volume} {278}},\ \bibinfo {pages} {20445} (\bibinfo {year}
  {2003})}\BibitemShut {NoStop}%
\bibitem [{\citenamefont {Iglesias}\ and\ \citenamefont
  {Devreotes}(2008)}]{iglesias2008}%
  \BibitemOpen
  \bibfield  {author} {\bibinfo {author} {\bibfnamefont {P.~A.}\ \bibnamefont
  {Iglesias}}\ and\ \bibinfo {author} {\bibfnamefont {P.~N.}\ \bibnamefont
  {Devreotes}},\ }\bibfield  {title} {\bibinfo {title} {Navigating through
  models of chemotaxis},\ }\href {https://doi.org/10.1016/j.ceb.2007.11.011}
  {\bibfield  {journal} {\bibinfo  {journal} {Curr. Opin. Cell Biol.}\ }\textbf
  {\bibinfo {volume} {20}},\ \bibinfo {pages} {35} (\bibinfo {year}
  {2008})}\BibitemShut {NoStop}%
\end{thebibliography}%

\end{document}